%% file: muscle.tex
\documentclass[sigconf,authorversion]{acmart}

\input{Sections/preamble.tex}
\input{Sections/template_args.tex}

\begin{document}


\title{MuscleVAE: Model-Based Controllers of Muscle-Actuated Characters}

\input{Sections/authors.tex}
\input{Sections/abstract.tex}
\input{Sections/article_info.tex}

\begin{teaserfigure}
  \centering
  \includegraphics[width=1\textwidth]{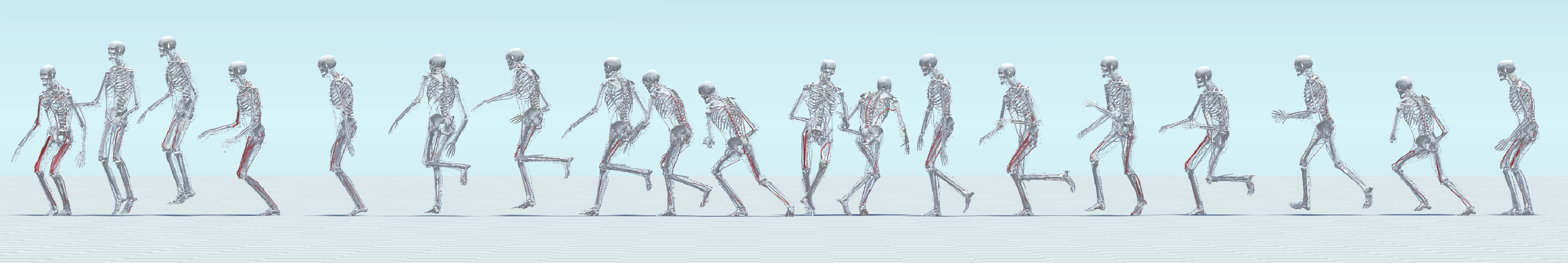}
  \caption{A simulated muscle-actuated character runs, walks, turns around, and performs single-leg and two-leg jumps. We present a model-based framework that uses variational autoencoders (VAE) for muscle control policy, thereby generating biologically plausible character motions. The color of the muscle curve represents the activation level; the redder the muscle, the greater its activation.}
  \label{fig:teaser}
\end{teaserfigure}

\maketitle

\input{Sections/introduction.tex}

\input{Sections/related_work.tex}

\input{Sections/muscle_system.tex}

\input{Sections/musclevae.tex}

\input{Sections/experiments.tex}

\input{Sections/conclusions.tex}

\input{Sections/acks.tex}

\bibliographystyle{ACM-Reference-Format}
\bibliography{muscle}

\input{Sections/x_figure_only_page.tex}

\input{Sections/supplimentary.tex}

\end{document}

%% file: Sections/preamble.tex
\usepackage{bm}
\usepackage{bbm}

\usepackage{graphicx}
\usepackage{tabularx}
\usepackage{subcaption}
\usepackage{multirow}
\usepackage{amsmath}
\usepackage{mathrsfs}

\newcommand{\vect}[1]{\bm{#1}}

\newcommand{\eqword}[1]{{\text{#1}}}

\usepackage{booktabs} 
\usepackage[ruled]{algorithm2e}

\usepackage{xspace}

\newcommand{\keep}[1]{}
\newcommand{\old}[1]{}

\newcommand{\loss}{\ensuremath{\mathcal{L}}}

\newcommand{\dataset}{\ensuremath{\mathcal{D}}}
\newcommand{\simBuffer}{\ensuremath{\mathcal{B}}}

\newcommand{\mvae}{MuscleVAE}

\makeatletter
\newcommand*\bigcdot{\mathpalette\bigcdot@{.5}}
\newcommand*\bigcdot@[2]{\mathbin{\vcenter{\hbox{\scalebox{#2}{$\m@th#1\bullet$}}}}}
\makeatother

\usepackage{url}
\usepackage{hyperref}
\newcommand{\fig}{Figure{}~}
\newcommand{\eqn}{Equation{}~}
\newcommand{\Tab}{Table{}~}

\newcommand{\stt}{\ensuremath{\vect{s}}}

\newcommand{\act}{\ensuremath{\vect{a}}}
\newcommand{\atv}{\ensuremath{\alpha}}
\newcommand{\dt}{\ensuremath{{\Delta t}}}

\newcommand{\world}{\ensuremath{{\omega}}}
\newcommand{\policy}{\ensuremath{{\pi}}}
\newcommand{\latent}{\ensuremath{\vect{z}}}
\newcommand{\Latent}{\ensuremath{\mathcal{Z}}}
\newcommand{\task}{\ensuremath{\vect{g}}}
\newcommand{\ml}{\ensuremath{l_{\mathrm{M}}}}

\newcommand{\mldot}{\ensuremath{\dot{l}_{\mathrm{M}}}}
\newcommand{\mltilde}{\ensuremath{\tilde{l}_{\mathrm{M}}}}

\newcommand{\mlt}{\ensuremath{l_{\mathrm{M}}^{\eqword{tpose}}}}
\newcommand{\kp}{\ensuremath{{k}_\eqword{p}}}
\newcommand{\kd}{\ensuremath{{k}_\eqword{d}}}
\newcommand{\force}{\ensuremath{{f}}}
\newcommand{\forcerm}{\ensuremath{\vect{f}}}
\newcommand{\point}{\ensuremath{\vect{p}}}
\newcommand{\targetload}{\ensuremath{u}}

%% file: Sections/template_args.tex
\AtBeginDocument{%
  \providecommand\BibTeX{{%
    \normalfont B\kern-0.5em{\scshape i\kern-0.25em b}\kern-0.8em\TeX}}}

\copyrightyear{2023}
\acmYear{2023}
\setcopyright{acmlicensed}
\acmConference[SA Conference Papers '23]{SIGGRAPH Asia 2023 Conference Papers}{December 12--15, 2023}{Sydney, NSW, Australia}
\acmBooktitle{SIGGRAPH Asia 2023 Conference Papers (SA Conference Papers '23), December 12--15, 2023, Sydney, NSW, Australia}
\acmPrice{15.00}
\acmDOI{10.1145/3610548.3618137}
\acmISBN{979-8-4007-0315-7/23/12}

\acmSubmissionID{178}



%% file: Sections/authors.tex
\author{Yusen Feng}
\email{ysfeng@stu.pku.edu.cn}
\orcid{0009-0008-8094-3610}
\affiliation{%
  \institution{Peking University}
  \streetaddress{No.5 Yiheyuan Road, Haidian District}
  \city{Beijing}
  \country{China}
  \postcode{100871}
}

\author{Xiyan Xu}
\email{xiyan_xu@pku.edu.cn}
\orcid{0009-0009-4609-064X}
\affiliation{%
  \institution{Peking University}
  \streetaddress{No.5 Yiheyuan Road, Haidian District}
  \city{Beijing}
  \country{China}
  \postcode{100871}
}

\author{Libin Liu}
\authornote{corresponding author}
\email{libin.liu@pku.edu.cn}
\orcid{0000-0003-2280-6817}
\affiliation{%
  \institution{Peking University}
  \city{Beijing}
  \country{China}
}
\affiliation{%
  \institution{National Key Lab of General AI}
  \city{Beijing}
  \country{China}
}
\renewcommand{\shortauthors}{Feng, Xu, Liu}

%% file: Sections/abstract.tex
\begin{abstract}
In this paper, we present a simulation and control framework for generating biomechanically plausible motion for muscle-actuated characters. We incorporate a fatigue dynamics model, the 3CC-r model, into the widely-adopted Hill-type muscle model to simulate the development and recovery of fatigue in muscles, which creates a natural evolution of motion style caused by the accumulation of fatigue from prolonged activities. To address the challenging problem of controlling a musculoskeletal system with high degrees of freedom, we propose a novel muscle-space control strategy based on PD control. Our simulation and control framework facilitates the training of a generative model for muscle-based motion control, which we refer to as MuscleVAE. By leveraging the variational autoencoders (VAEs), MuscleVAE is capable of learning a rich and flexible latent representation of skills from a large unstructured motion dataset, encoding not only motion features but also muscle control and fatigue properties. We demonstrate that the MuscleVAE model can be efficiently trained using a model-based approach, resulting in the production of high-fidelity motions and enabling a variety of downstream tasks.
\end{abstract}

%% file: Sections/article_info.tex
\ccsdesc[500]{Computing methodologies~Animation}
\ccsdesc[500]{Computing methodologies~Physical simulation}
\ccsdesc[300]{Computing methodologies~Reinforcement learning}
\ccsdesc[300]{Computing methodologies~Neural networks}

\keywords{muscle, fatigue simulation, motion control, generative models, VAE}

%% file: Sections/introduction.tex
\section{Introduction}

Animating characters using detailed musculoskeletal models offers the potential for highly realistic and accurate character movements. However, progress in muscle-actuated character animation has been comparatively slow relative to advancements in rigid body character animation. This lag is primarily attributed to the challenges associated with the high-dimensional nature of muscle actuation space, which leads to substantial simulation times and poses significant training challenges. Additionally, muscle dynamics models used in recent computer animation research are often overly simplified. Some important factors, such as the effects of fatigue cumulation, remain inadequately explored.

In this paper, we present a comprehensive simulation and learning framework for muscle-actuated characters. Building upon the widely-adopted Hill-type muscle models~\cite{hill1938heat,zajac1989muscle}, we develop a new control scheme in the muscle space that efficiently determines muscle forces while taking into account the constraints imposed by the musculotendon models. Additionally, we incorporate fatigue effects into our muscle simulator to generate motions that are more biologically accurate. Based on this simulation framework, we utilize a model based on Variational Autoencoder (VAE) to learn skill embeddings from unorganized motion data. This learned latent space of motion skills encompasses both muscle functionality and coordination, making it applicable to various downstream tasks.

To mitigate the difficulties caused by the high degrees of freedom of the muscle system, several recent successful animation systems \cite{ryu2021functionality, park2022generative, jehee19muscle} employ a two-level control framework proposed by \cite{jehee19muscle}. This framework learns joint-level PD control using reinforcement learning and trains a specialized network to coordinate muscles to realize the computed joint torques. However, relying on joint-level control as the driving signal may not accurately capture the biomechanical patterns of muscle activation and may be susceptible to overfitting in specific torque regions. 
In this paper, we opt for controlling the character directly in the muscle space. 
We attach a PD servo to each muscle fiber and make the control policy compute the target length for these PD servos. The resulting muscle forces are then confined to the range determined by musculotendon models and fatigue dynamics. We find that such a straightforward strategy can effectively facilitate the learning of complex skills and can be easily incorporated into the training framework.

Fatigue is a common phenomenon in real muscle actuation systems. However, due to the challenges associated with acquiring fatigue data, quantitative studies on this issue are often limited to simple movements and specific fatigue parameters of certain body parts. Character animation studies examining the effects of fatigue on motion are also sparse \cite{cheema2020predicting,komura2000creating}. In this paper, we consider fatigue as an integral part of a complete musculoskeletal system. We incorporate the 3CC-r muscle fatigue model \cite{looft20183ccr} into our system, expanding its scope to encompass full-body movements. Our control policy automatically shifts its strategy at different fatigue levels, generating natural change of motion patterns that are biomechanically plausible.

To learn a versatile skill representation from a large, unorganized motion dataset, many recent studies rely on model-free reinforcement techniques combined with adversarial networks \cite{peng2022ase} or autoencoders \cite{won2022physicsvae}. However, model-free approaches can suffer from sample efficiency issues and can be difficult to converge on high-dimensional problems. Recently, model-based approaches have proven to be data efficient and stable in training complex motion controllers \cite{supertrack2021,controlvae2022,hafner2023mastering}. In these approaches, a world model is learned to capture the complex dynamics of character motion, which allows for differentiable training objectives. In this paper, we adopt a model-based method, ControlVAE \cite{controlvae2022}, to learn generative control policies for our muscle-actuated characters. We integrate our differentiable muscle-space controller into the world model, utilizing gradient information to facilitate the learning of muscle activation coordination.

In summary, our work makes two principal contributions: (1)~We propose a novel simulation and control framework for muscle-actuated characters. This framework incorporates fatigue effects in simulation, and our muscle-space control mechanism facilitates biologically plausible control of complex human motions; (2)~We develop a generative control policy for muscle-driven characters.  Trained using a model-based approach, this policy not only provides a versatile skill representation for numerous downstream tasks but also automatically adjusts to fatigue levels. This adaptability ensures that characters exhibit natural variations in motion patterns during extended activities.

%% file: Sections/related_work.tex
\section{Related works}

\subsection{Muscle Modeling and Simulation}

Muscle modeling and simulation has been a long-standing topic in both biomechanics and computer graphics. The Hill-type model, proposed by \cite{hill1938heat} and expanded by \cite{zajac1989muscle}, numerically models musculotendon dynamics. OpenSim \cite{delp2007opensim, seth2018opensim} leverages this model to simulate human body movement. \citet{wang2022differentiable} accounted for muscle inertia, creating a compatible framework for the Hill-type muscle. For volumetric simulation, finite element methods (FEM) are used to model muscular soft tissue deformation \cite{zhu1998real, lee2009comprehensive, fan2014active}. EMU \cite{modi2021emu} handles heterogeneously stiff meshes with better efficiency than FEM. Recently, various frameworks and suites for muscle simulation have been proposed, including open-source ones such as \cite{todorov2012mujoco, MyoSuite2022}, as well as commercial ones such as \cite{Geijtenbeek2021Hyfydy}. Various specific muscle models for body parts such as the face, neck, shoulder, and hand \cite{ichim2017phace, srinivasan2021learning,yang2022implicit, lee2006heads, van1994finite, maurel2000human, sueda2008musculotendon, li2022nimble}, 
and animals like ostrich \cite{labarbera2022ostrichrl}  have also been developed.

Muscle fatigue, which is performance degradation resulting from intense muscle exercise, has been a research topic for years in biomechanics and related fields. \citet{giat1993musculotendon} analyzed fatigued quadriceps muscle, presenting a fatigue-recovery model \cite{giat1993musculotendon, giat1996model}. With validations on calculated METs (Maximum Endurance Times), \citet{ma2009new} modeled fatigue patterns. \citet{musclemodelfatigue} proposed a framework that considers the fatigue-related changes in motor unit force, but their model does include recovery from fatigue. The works of \cite{komura2000creating, liu2002dynamical} estimated tired poses and quantified muscle fatigue and recovery. The three-compartment controller fatigue model (3CC), proposed by \citet{xia2008theoretical}, predicts muscle fatigue in complex movements. The 3CC model was improved by adding a rest recovery parameter \cite{looft20183ccr} and was integrated into reinforcement learning (RL) reward by \citet{cheema2020predicting} to design policies for mid-air interaction movements. We build our system on the Hill-type muscle model for its simplicity and efficiency. We augment this model with a modified 3CC-r model to simulate fatigue, and a PD control mechanism to allow efficient training of complex skills.

\subsection{Motion Control and DRL}

Reproducing realistic, interactive motions in physics-based simulation is a challenging problem. Early works relied on human insights and designed torque-actuation control strategies \cite{hodgins1995animating}, while later works used learning algorithms, optimal control, and abstract models \cite{sok2007simulating, muico2011composite, yin2007simbicon}. Recently, deep reinforcement learning (DRL) has shown potential in various tasks \cite{Liu2017_Learning, peng2018deepmimic, yin2021discovering, won2021control, lee2021learning}. Beyond merely tracking motion trajectory, a number of recent studies successfully learned generative control policies that allow efficient accomplishment of downstream tasks \cite{peng2022ase,won2022physicsvae,controlvae2022}.

Muscle-actuated control, built upon detailed musculoskeletal dynamics, has the potential to synthesize more realistic human postures compared to those produced by joint-actuated control \cite{komura2000creating}. Numerous efforts have been made to establish robust muscle-actuated control systems for tasks such as locomotion, swimming, and hand manipulation \cite{wang2012optimizing,geijtenbeek2013flexible,lee2014locomotion,geyer2010muscle,si2014realistic,tsang2005helping}. Combined with deep reinforcement learning, several recent successful frameworks \cite{ryu2021functionality, park2022generative, jehee19muscle} achieve tracking control of various motion skills within a uniform framework. To mitigate the challenges posed by the high degrees of freedom of the muscle system, \citet{jehee19muscle} employ a two-level imitation learning algorithm, where a joint-level PD control policy is combined with a separate network to coordinate muscles and realize the computed joint torques. A more recent work, DEP-RL \cite{schumacher2023:deprl}, also shows that this problem can be partially addressed by employing better exploration techniques in reinforcement learning.
Our framework also utilizes deep reinforcement learning to train complex control policies. We directly compute muscle actuation using a novel muscle-space PD control mechanism, eliminating the need for guidance from a joint-level controller. We also employ a model-based reinforcement method, ControlVAE \cite{controlvae2022}, to effectively train our control policies. Here, muscle states and fatigue information are taken into account, enabling the policy to adapt under different conditions.

%% file: Sections/muscle_system.tex
\section{Muscle System}
\label{muscle_sys}

\subsection{Muscle Modeling}
\label{muscle_model}

\begin{figure}
  \includegraphics[width=0.23\textwidth]{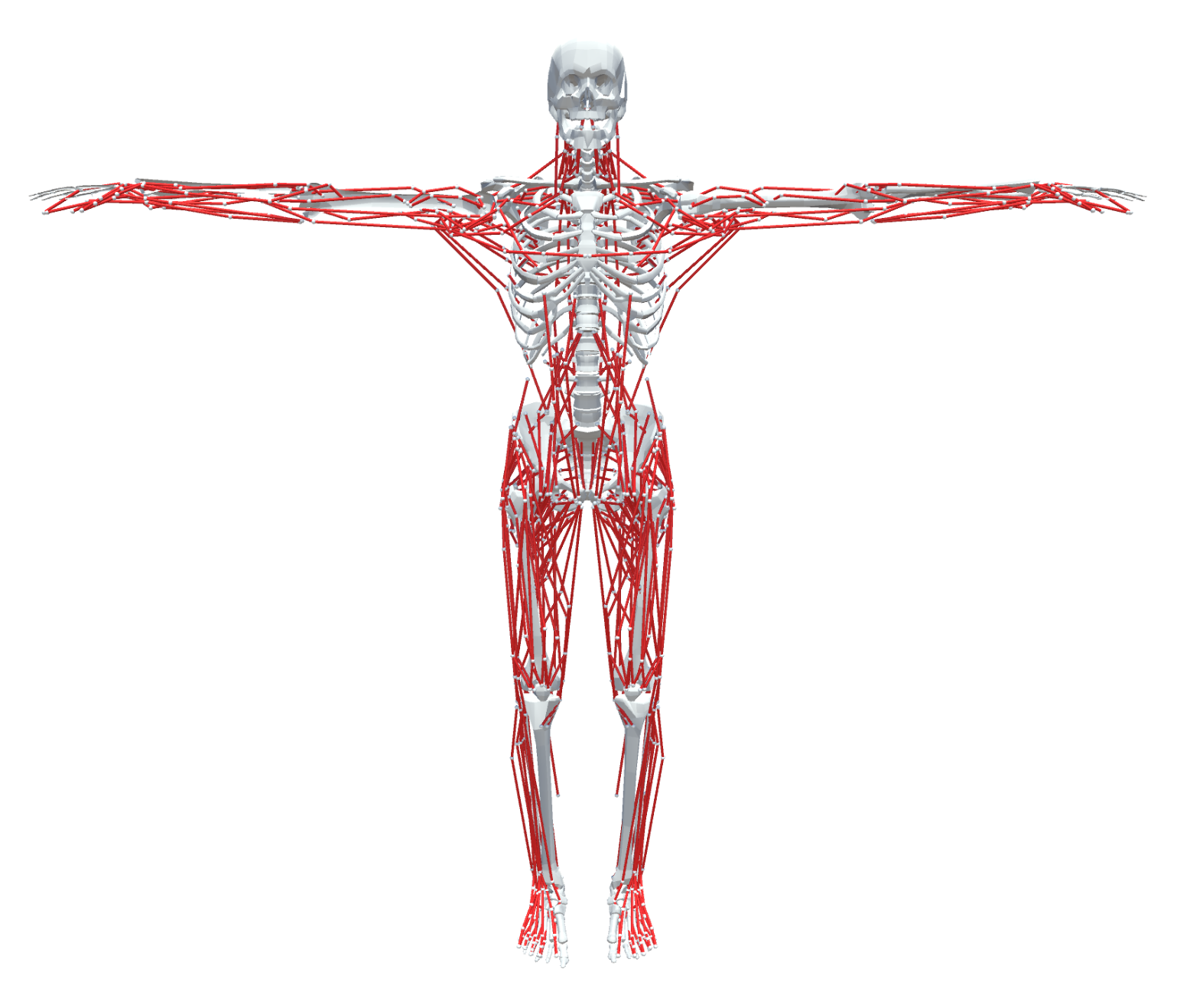}
  \includegraphics[width=0.23\textwidth]{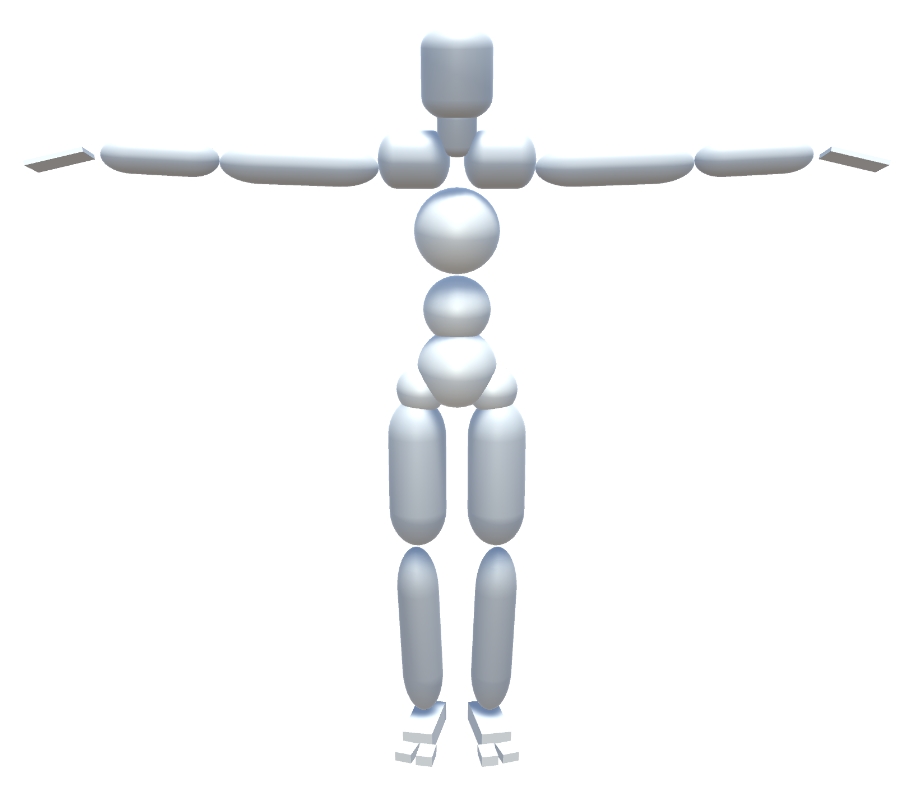}
  \caption{The muscle model and the simulated rigid bodies. Muscles are colored in red to highlight the connections between muscles and bones.}
  \label{fig:muscle_model}
\end{figure}

Our simulated character is adapted from the musculoskeletal model developed by \citet{jehee19muscle}, with minor modifications made to enforce symmetry between the left and right sides of the character. As shown in Figure \ref{fig:muscle_model}, the character model consists of eight revolute joints and fourteen ball-and-socket joints. It is actuated by 284 muscles. These muscles are the sole drivers of motion, while the joints provide the necessary physical constraints.

Muscles attach to bones via tendons at their ends, known as the origin and insertion. Following~\cite{jehee19muscle,delp2007opensim}, we use a simplified muscle model, where each muscle is represented as polylines and may span across multiple joints. These polylines are defined by a set of anchor points, and the muscle force is considered to be transferred to the bones through these anchor points. When the character moves, the placement of these anchors is computed using Linear Blend Skinning (LBS).

\subsection{Muscle Dynamics}
\label{muscle_dynamics}

Muscles are often modeled using a simplified, three-element structure known as the Hill-type muscle model \cite{hill1938heat, zajac1989muscle}. This model comprises a contractile element (CE), a parallel elastic element (PE), and a tendon element. The CE represents the muscle fibers, which contract based on the muscle's activation state. It generates an active contractile force that is proportional to the level of activation $\atv \in [0, 1]$. The PE represents  the passive elastic material surrounding the muscle fibers and produces a passive, non-linear spring force. Following the common practice in previous works \cite{jiang2019synthesis, jehee19muscle, geijtenbeek2013flexible}, we further simplify this model by neglecting changes in tendon length, assuming zero pennation angles, and calculating muscle length using the polylines. The muscle force generated by each muscle is then determined by the forces from the CE and PE as
\begin{equation}
    f_{\text{m}} = \atv f_{\text{CE}}^l(\bar{l})f_{\text{CE}}^v(\dot{\bar{l}})+f_{\text{PE}}(\bar{l}),
\label{equ:hill}
\end{equation}
where $\bar{l}$ and $\dot{\bar{l}}$ represent the normalized muscle length and its rate of change, respectively. We compute $\dot{\bar{l}}$ using the finite difference between consecutive frames.
$f_{\text{CE}}^l$, $f_{\text{CE}}^v$, and $f_{\text{PE}}$ are the active force-length function, the force-velocity function, and the passive force-length function, respectively. The exact form of these functions is experimentally determined and can be found in the supplementary document.

\begin{figure}
    \includegraphics[width=0.3\textwidth]{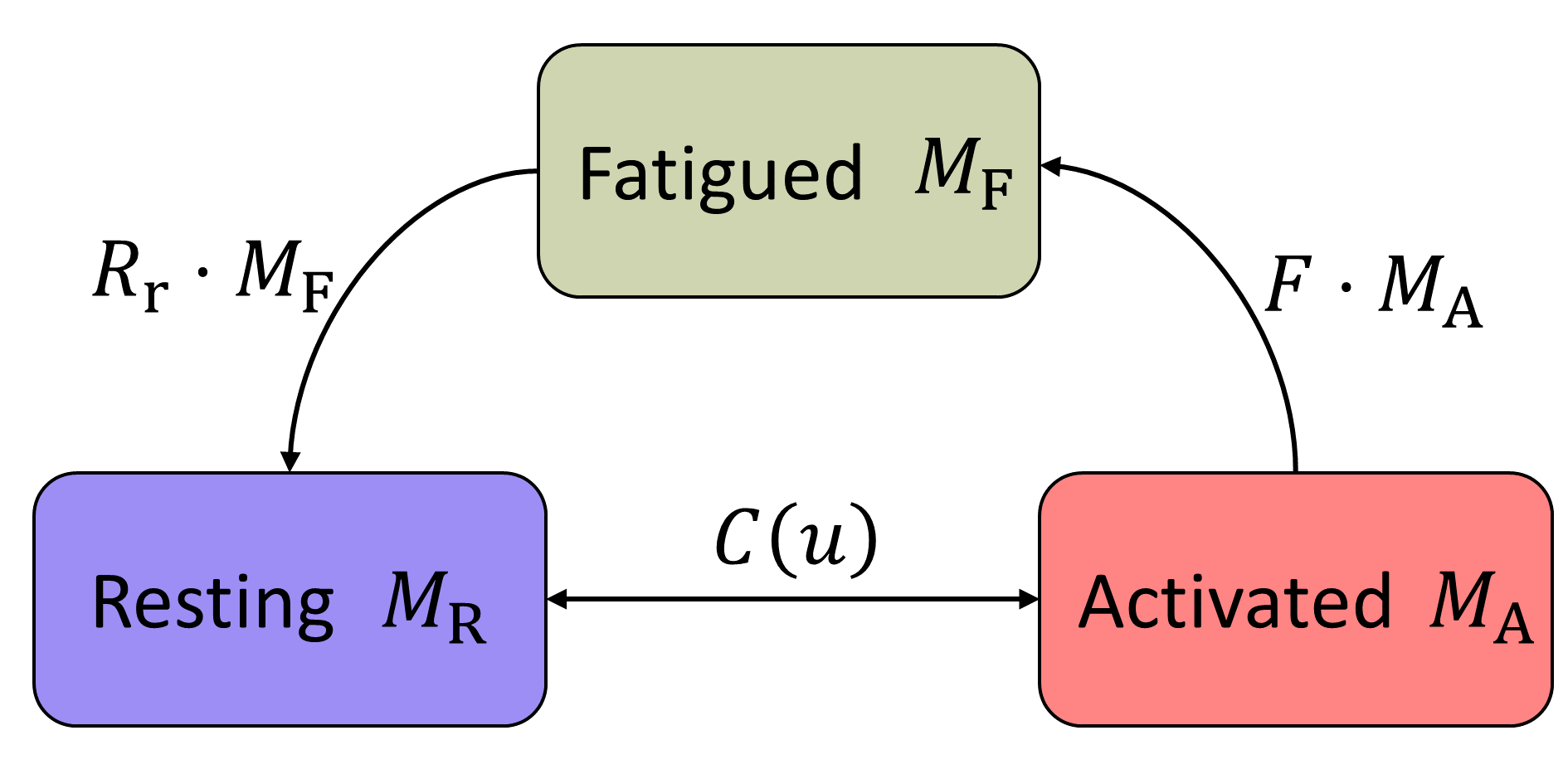}
    \caption{The 3CC-r model assumes that muscle actuators can be in one of three possible compartments. The differential quantities within these three compartments evolve according to their respective relationships.}
    \label{fig:3ccr_model_0}
\end{figure}

\subsection{Fatigue Dynamics}

Fatigue is the phenomenon in which a particular muscle cannot maintain the required force due to the accumulation of substances that cause fatigue. This loss of strength in specific muscles leads to a redistribution of muscle activation and a consequent alteration of movement patterns. 
In this paper, we adopt the 3CC-r model \cite{looft20183ccr} to simulate fatigue effects. This model is an enhanced version of the Three-Compartment Controller (3CC) model \cite{xia2008theoretical}, incorporating additional factors for improved alignment with experimental data. The 3CC-r model assumes that each muscle consists of multiple hypothetical muscle-tendon actuators. Each of these actuators is presumed to be in one of three possible states: \emph{Activated} ($M_{\text{A}}$), \emph{Resting} ($M_{\text{R}}$), and \emph{Fatigued}~($M_{\text{F}}$). Each $M_{*}$ here represents the percentage of actuators in a specific state. Actuators in the \emph{Activated} state are considered to be ideally activated, producing maximal contractile forces. In contrast, actuators in both the \emph{Resting} and \emph{Fatigued} states generate no contractile forces. Based on these assumptions, $M_{\text{A}}$ can be viewed as representing the activation level of the entire muscle, corresponding to $\atv$ in Equation~\eqref{equ:hill}.

In the 3CC-r model, the actuators in the \emph{Resting} and \emph{Activated} states can transition to the other state when needed. Once activated, the \emph{Activated} actuators become \emph{Fatigued} over time at a given rate, and the \emph{Fatigued} actuators can recover gradually and revert to the \emph{Resting} state. Figure~\ref{fig:3ccr_model_0} illustrates this process. The values of $M_{\text{A}}$, $M_{\text{R}}$, and $M_{\text{F}}$ are governed by a set of differential equations:
\begin{gather}
      \dot{M}_\text{A}=C(\targetload)-F \cdot M_\text{A}  \label{equ:3cc-MA}\\
      \dot{M}_\text{R}=-C(\targetload)+R_{\text{r}} \cdot M_\text{F} \qquad
      \dot{M}_\text{F}=F \cdot M_\text{A}-R_{\text{r}} \cdot M_\text{F}        
    \label{equ:3cc-0},
\end{gather}
where $F$ and $R_{\text{r}}$ denote the fatigue and recovery coefficients, respectively. The function $C(\targetload)$ denotes the transfer rate between $M_{\text{R}}$ and $M_{\text{A}}$, determined by the difference between the target load, $\targetload$, and the current activation level $M_{\text{A}}$. The target load, $\targetload\in[0,100\%]$, describes a desired level of muscle activation that the character's brain wishes to use to perform a motion. 
The effect of $C(\targetload)$ is akin to the activation dynamics~\cite{delp2007opensim}, where $\targetload$ and $M_{\text{A}}$ corresponds to the excitation and activation signals, respectively. 
Following \cite{xia2008theoretical,looft20183ccr}, we use a piecewise function that increases monotonically with~$u$ to formulate $C(\targetload)$. We refer readers the supplementary materials for its accurate form.

\subsection{Muscle Space Control}
\label{sec:muscle_actor}
The Hill-type model generates muscle forces based on the activation signals. However, in our early experiments, we found that using muscle activation levels as action space can lead to poor convergence. To remedy this problem, we use a PD control-like formulation at the muscle level to calculate the force applied by the muscle. 
The PD muscle force operates as
\begin{equation}
\force_{\text{pd}} = \max\left({0}, \kp\cdot(\mltilde-\ml)-\kd\cdot\mldot\right) ,
\label{equ:pd_force}
\end{equation}
where $\mltilde$ denotes a target muscle length, 
$\ml$ and $\mldot$ refer to the current length of the muscle and its rate of change, respectively. $k_p$ and $k_d$ are predefined PD gains. Since the force generated by the muscle can only be contractile, any negative component is eliminated.

We can compute the muscle activation $\atv_{\text{pd}}$ that leads to the PD muscle force $\force_{\text{pd}}$ using
\begin{equation}
  \atv_{\text{pd}} = \frac{\force_{\text{pd}} - f_{\text{PE}}(\bar{l})}{f_{\text{CE}}^l(\bar{l})f_{\text{CE}}^v(\dot{\bar{l}})} \quad.
  \label{equ:alpha_pd}
\end{equation}
However, $\atv_{\text{pd}}$ may not be achievable due to the muscle constraints and fatigue.
In the Hill-type model, the muscle activation $\atv$ is confined to the range $[0, 1]$ and evolves according to the fatigue dynamics of Equation~\eqref{equ:3cc-MA}. As a result, the PD muscle force computed above is not always realizable. We argue that the feasible range of muscle force can be defined by a set of upper and lower bounds $[f_{\eqword{lb}}, f_{\eqword{ub}}]$. The final force applied to the character is then computed by
\begin{equation}
    \force = \text{clip}(\force_{\text{pd}}, \; f_{\eqword{lb}}, \; f_{\eqword{ub}}).
    \label{equ:force}
\end{equation}
To find $f_{\eqword{lb}}$ and $f_{\eqword{ub}}$, considering that the muscle activation, $\atv$, and percentage of activated actuators, $M_{\text{A}}$, are equivalent, the discrete form of the fatigue dynamics of Equation~\eqref{equ:3cc-MA} can be written as

\begin{gather}
    ({\tilde{\atv}-\atv})/{\dt{}}=C(\targetload)-F \atv,
    \\    
    \text{or,}\quad \tilde{\atv}=\tilde{\atv}(\targetload)=(1-\dt{}F)\atv + \dt{}C(\targetload),
    \label{equ:3cc-MA-discrete} 
\end{gather}
where $\dt{}$ represents the time interval. Equation~\eqref{equ:3cc-MA-discrete} suggests that the next activation level, $\tilde{\atv}$, is determined by the previous muscle activation level, $\atv$, and the target load, $\targetload$. Given that $\targetload$ can be freely selected within the range $[0,1]$ and considering that $C(\targetload)$ increases monotonically with $\targetload$, it follows that $\tilde{\atv}\in[\tilde{\atv}(\targetload=0),\tilde{\atv}(\targetload=1)]$. Furthermore, based on the Hill-type model presented in Equation~\eqref{equ:hill}, the muscle force also rises monotonically with muscle activation. Thus, the force bounds can be computed as

\begin{align}
    f_{\eqword{lb}} &= \tilde{\atv}(\targetload=0) \cdot f_{\text{CE}}^l(\bar{l})f_{\text{CE}}^v(\dot{\bar{l}})+f_{\text{PE}}(\bar{l}) \\
    f_{\eqword{ub}} &= \tilde{\atv}(\targetload=1) \cdot f_{\text{CE}}^l(\bar{l})f_{\text{CE}}^v(\dot{\bar{l}})+f_{\text{PE}}(\bar{l})
\end{align}

We employ a control policy $\pi$ to calculate an action vector $\act$ that contains the action $a$ for each muscle. The target muscle length, $\mltilde$, is then computed using 
\begin{equation}
    \mltilde = (a + 1.0) \cdot \mlt,
    \label{equ:action_to_target_length}
\end{equation}
where $\mlt$ denotes the reference muscle length, computed in the T-pose of the character. After computing $\force_{\eqword{pd}}$, we apply the force bounds using Equation~\eqref{equ:force} and solve for the target load $\targetload$ that yields $\force$. Finally, $\force$ is applied to actuate the character, while the corresponding $\targetload$ is used to simulate muscle fatigue using the 3CC-r model.

We find that this muscle PD control mechanism significantly facilitates training compared to using activation control. The improvement can be attributed to the introduction of local feedback loops by the PD control, which allows the system to self-adjust and self-correct. This finding aligns with the comparison made between joint-level PD control and torque control done by \cite{2017-SCA-action}. In this context, muscle activations can be analogously related to joint torques, and our PD muscle force resembles the joint-level PD control.

The procedures described above are all differentiable. We can concisely write them as
\begin{equation}
    \tilde{\bm{\atv}},\stt_{\text{muscle}} = D(\stt_{\text{skeleton}},\stt_{\text{fatigue}},{\act},\bm{\atv}). 
\end{equation}
In this formulation, we use bold symbols to collectively represent the corresponding values for all the muscles. $\tilde{\bm{\atv}}$ and $\bm{\atv}$ represent the next and previous activation values, respectively. $\stt_{\text{skeleton}}$ refers to the state of the skeleton, while $\stt_{\text{fatigue}}$ denotes the fatigue state of the muscles. The muscle kinematic state, $\stt_{\text{muscle}}$, contains the $\ml$ and $\mldot$ values and is derived from $\stt_{\text{skeleton}}$ using LBS. $\act$ represents the action computed by the policy $\pi$. And lastly, $D$ denotes the entire procedure.

%% file: Sections/musclevae.tex
\section{Muscle VAE}
\label{muscle_vae}

\begin{figure*}
    \centering
    \includegraphics[width=0.9\linewidth]{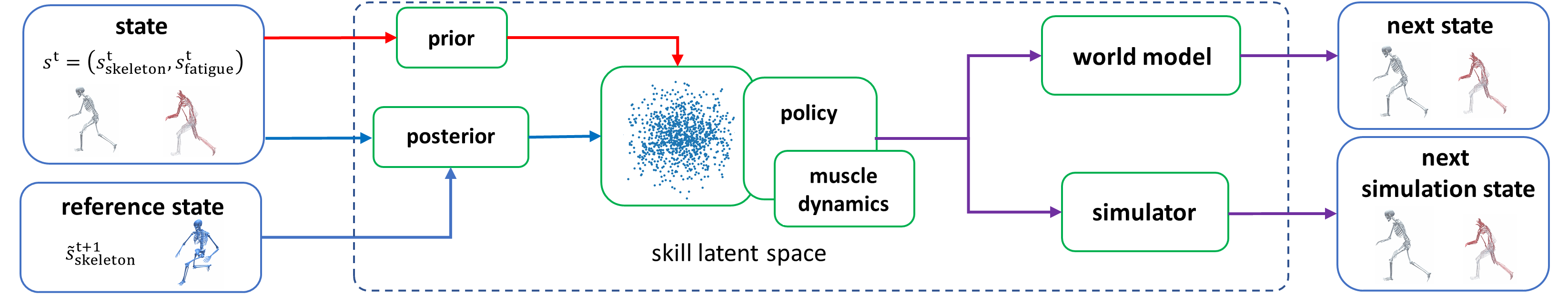}
    \caption{Overview of our \mvae{} System. }
    \label{Fig:Pipeline}
\end{figure*}

In this section, we introduce our model-based, muscle-actuated motion control framework, which we refer to as \emph{MuscleVAE}. This framework is inspired by ControlVAE \cite{controlvae2022}. Formally, as depicted in Figure~\ref{Fig:Pipeline}, our objective is to learn a policy, $\policy$, associated with a latent space, denoted as $\Latent$. The latent space $\Latent$ encapsulates all the skills of a motion dataset. When a latent code $\latent\in\Latent$ is selected from this space, the policy $\policy(\act|\stt, \latent)$ converts it into an action vector $\act$ according to the current state of the character $\stt$. This action vector $\act$ then determines muscle forces, as outlined in Sec~\ref{sec:muscle_actor}, that actuates the character to perform a particular skill.

The state $\stt$ represents the complete state of the character, comprising both the skeleton state $\stt_{\text{skeleton}}$ and the fatigue state $\stt_{\text{fatigue}}$. Note that we do not include the muscle state $\stt_{\text{muscle}}$ in $\stt$ as it can be directly computed from $\stt_{\text{skeleton}}$. We adopt the same skeleton state representation as described in \cite{controlvae2022}, which includes the positions, orientations, and velocities of all rigid bones. When provided with motion data as input, our framework extracts skeleton states $\{\tilde{\stt}_{\text{skeleton}}^t\}$ and muscle states $\{\tilde{\stt}_{\text{muscle}}^t\}$ from it, using these as the reference states. Here, $t$ represents the time index, the tilde symbols indicate quantities that are derived from the motion data, rather than from simulation, and the braces denote a sequence of such quantities. The input motions do not need fatigue data.

Following ControlVAE~\cite{controlvae2022}, we train a motion encoder, represented as a posterior distribution $q(\latent^t|\stt^t,\tilde{\stt}^{t+1}_{\text{skeleton}})$, to convert a motion into latent codes $\{\latent^t\}$, where $\{\tilde{\stt}^{t+1}_{\text{skeleton}}\}$ are the reference states extracted from the motion and $\{\stt^t\}$ are the simulation states of the character.  These latent codes $\{\latent^t\}$ can be decoded by the policy $\policy$ into muscle forces to actuate the character, allowing it to reproduce the input motion in simulation. The VAE framework imposes additional regularization on the posterior distribution~$q(\latent^t|\stt^t,\tilde{\stt}^{t+1}_{\text{skeleton}})$, encouraging it to stay close to a prior distribution~$p(\latent)$. This scheme ensures that a random latent code sampled from the prior distribution can be decoded into a valid motion. A common choice for the prior distribution $p(\latent)$ is the standard normal distribution $\mathcal{N}(0, 1)$. However, \citet{controlvae2022} suggest that a state-dependent prior distribution $p(\latent|\stt)$ can provide better performance. We thus adopt the same prior distribution in our framework. At runtime, the latent code can be either computed by the posterior encoder, sampled using the prior distribution, or provided by a high-level policy of a downstream task.

We formulate the components of the MuscleVAE, specifically the policy $\policy(\act|\stt,\latent)$, the posterior distribution $q(\latent^t|\stt^t,\tilde{\stt}^{t+1}_{\text{skeleton}})$, and the state-dependent prior distribution $p(\latent|\stt)$, as normal distributions in the form of $\mathcal{N}(\vect{\mu}_*(~\bigcdot~;\theta_*),\sigma_*^2\vect{I})$. Here, $\sigma_*$ is a predefined standard deviation, and the mean $\vect{\mu}_*(~\bigcdot~;\theta_*)$ is represented by a neural network with trainable parameters $\theta_*$. The structure of these networks can be found in the supplementary materials.

\subsection{Training}
We train MuscleVAE on a dataset containing multiple unstructured motion sequences, using an approach similar to ControlVAE~\cite{controlvae2022}. During the training process, our framework iteratively extracts short motion clips from the dataset, encodes them using the posterior distribution $q(\latent|\stt,\tilde{\stt}_{\text{skeleton}})$, decodes the resulting latent codes with the policy $\policy(\act|\stt,\latent)$, and reconstructs the motion clips via simulation. We train all components of MuscleVAE simultaneously, aiming to minimize the reconstruction error while ensuring the posterior distribution and the state-dependent prior distribution $p(\latent|\stt)$ stay close to each other. Formally, this objective can be written as
\begin{equation}
    \mathcal{L}_{\text{VAE}} = \mathcal{L}_{\text{rec}}  + \beta \mathcal{L}_{kl} +  \mathcal{L}_{\text{act}},
    \label{eqn:loss_VAE_0}
\end{equation}
where $\beta$ is a weight parameter suggested by \citet{Higgins2017betaVAELB}. 
The reconstruction loss $\mathcal{L}_{\text{rec}}$ measures the discrepancy between the simulated states and the reference states. In MuscleVAE, we consider both the skeletal motion and muscle length, so
\begin{align}
    \mathcal{L}_{\text{rec}} = \sum_{t=0} \gamma^t \left[\left\Vert \tilde{\stt}_{\text{skeleton}}^{t} - {\stt}_{\text{skeleton}}^{t} \right\Vert^2_{W} +
    \left\Vert \tilde{\stt}_{\text{muscle}}^{t} - {\stt}_{\text{muscle}}^{t} \right\Vert^2_{W'} \right], 
    \label{eqn:loss_recon}
\end{align}
This function is evaluated over the motion sequence. Here, $W$ and $W'$ represent weight matrices, and $\gamma$ is a discount factor. The KL-divergence loss 
\begin{align}
\mathcal{L}_\text{kl}=\sum_{t=0} \gamma^t \mathcal{D}_{\eqword{KL}}\left(q(\latent^t|\stt^t,\tilde{\stt}^{t+1}_{\text{skeleton}}) ~\Vert~ p(\latent^t|\stt^t)\right)
\label{eqn:kl_loss}
\end{align}
penalizes the difference between the prior and posterior distributions. {Finally, to mitigate excessive control and satisfy the biological requirements of minimal bioenergy and activation thresholds, we use a combination of $L_1$ and $L_2$ losses on the activation level, hence the regularization loss $\mathcal{L}_{\text{act}}$ is defined as}
\begin{align}
    \mathcal{L}_{\text{act}} =  \sum_{t=0} \gamma^t \left(w_{a_1}\left\|\atv_{\text{pd}}^t \right\|_1+w_{a_2}\left\|\atv_{\text{pd}}^t \right\|_2\right),
\end{align}
{where $w_{a_1}$ and $w_{a_2}$ are weights of the regularization terms}.

\subsubsection{Model-based Learning}
The objective function, as shown in Equation~\eqref{eqn:loss_VAE_0}, cannot be directly optimized since evaluating it requires going through a complex rigid body simulation. Our system treats the simulation procedure as a black box. This approach potentially allows our system to accommodate various simulation backends, but it also makes the simulation procedure non-differentiable. Instead, as suggested by \citet{controlvae2022}, we adopt a model-based training procedure for our MuscleVAE, given its proven efficiency.

Briefly speaking, we train a world model, $\world$, to approximate the dynamics of the musculoskeletal system, $\stt^{t+1} = \world(\stt^t, \act^t)$. Then, $\world$ is used as a substitute for the real simulation when evaluating Equation~\eqref{eqn:loss_VAE_0}. We formulate $\world$ as a neural network, making it differentiable and enabling the optimization of Equation~\eqref{eqn:loss_VAE_0} through gradient-based methods.
To train $\world$, our system first generates a simulation sequence $\{\stt^0,\act^0,\stt^1,\act^1,\dots\}$ by repeatedly executing the current MuscleVAE to a track random motion sequence in the real simulation. Then, starting from $\stt^0$, our system creates a synthetic sequence by executing the same series of actions $\{\act^t\}$ in the world model. This results in a sequence of synthetic states $\{\bar{\stt}^t\}$, where $\bar{\stt}^{t+1}=\world(\bar{\stt}^t, \act^t)$. At last, $\world$ is trained by optimizing the objective 
\begin{align}
    \mathcal{L}_{w} = \sum_{t=0} \left\Vert {\stt}^{t+1} - \bar{\stt}^{t+1} \right\Vert^2_{\tilde{W}} 
    \label{eqn:loss_world_model_0}
\end{align}
against these simulation samples. $\tilde{W}$ represents a weight matrix. The training of the world model and that of MuscleVAE's components are interleaved. When one is being trained, the other is frozen.

We employ a world model similar to that used in ControlVAE~\cite{controlvae2022}, which is formulated in maximal coordinates. In practice, we find that our musculoskeletal system is more sensitive to the accumulative error of the world model than the rigid body system used by \citet{controlvae2022}, especially in the early stage of the training. This is because a small change in the length of certain muscles can lead to excessive muscle forces. We thus employ an additional differentiable forward kinematics pass to mitigate such errors. 

During training,  we randomly switch between reference motion clips to expose the model to different motion patterns. This abrupt switching also helps the model adaptively learn from mismatched poses and velocities and recover from such disparities. Additionally, we randomly set fatigue states for the character at the beginning of each training rollout to allow the trained MuscleVAE to be robust in different fatigue configurations. For the formulation of the world model and the training details, we refer readers to \cite{controlvae2022} and the supplementary materials.

\subsubsection{High-Level Policies}
With a trained MuscleVAE, our system further enables a downstream task to be accomplished through a goal-conditioned task policy $\policy(\latent|\stt,\task)$. This task policy $\policy(\latent|\stt,\task)$ operates in the latent space $\Latent$ and outputs latent codes based on the current state $\stt$ and the goal $\task$ of the task. Following \cite{controlvae2022}, we train $\policy(\latent|\stt,\task)$ in a model-based manner with the trained MuscleVAE and the world model kept fixed during this phase. When given the objective function of the task, denoted as $\mathcal{L}_{\task}$, our system repeatedly executes the task policy $\policy(\latent|\stt,\task)$ and then decodes the resulting latent codes into a motion sequence using both the policy $\policy$ and the world model $\world$. The performance of these motion sequences is evaluated against $\mathcal{L}_{\task}$, and the policy $\policy(\latent|\stt,\task)$ is subsequently updated to minimize $\mathcal{L}_{\task}$.

%% file: Sections/experiments.tex
\section{Experiments}

\subsection{System Setup}
The muscle-actuated character depicted in Figure~\ref{fig:muscle_model} is used in all our experiments. It has a height $1.68$\,m, weighs $61.4$\,kg, consists of 23 rigid bodies connected by 22 joints, and is actuated by 284 muscles. The muscle model, including both the muscle dynamics and fatigue dynamics, operates at a frequency of $120$\,Hz. The simulation framework is implemented based on the rigid body simulator, Open Dynamics Engine~(ODE). To ensure numerical stability with this relatively large time step, we additionally incorporate implicit joint damping into ODE, as suggested by \citet{Liu2013_Simulation}. The \mvae{} is implemented and trained using PyTorch~\cite{pytorch2019} and runs at a lower frequency of 20\,Hz. When the \mvae{} policy computes an action, this action is reused for the subsequent six simulation steps. The entire system achieves real-time performance, enabling interactive control of the simulated character. 

We train our \mvae{} on a dataset containing approximately 25 minutes of motion sequences. These motion sequences are selected from the open-source LaFAN dataset \cite{harvey2020lafan}. They include a variety of locomotion skills, including walking, running, turning, hopping, and jumping.
The model is trained for 20,000 iterations. Our unoptimzed implementation takes about 1 week to train using six parallel threads on a workstation equipped with Intel Xeon Gold 6133 CPU and a single NVIDIA GTX 3090 graphics card.

\subsection{Evaluation}
We evaluate the effectiveness of our learned \mvae{} using three types of tasks: tracking, generation, and fatigue simulation. 

\paragraph{Tracking} 
The \mvae{} trained on the large locomotion dataset can be used to track other similar locomotion clips, even if they were not used for training. In this task, the posterior distribution is used to compute the latent codes of the input motion. The character can perform the input motion accurately over an extended time frame. When given another clip, it can figure out a smooth transition and then tracks the new target motion. If the two clips differ too significantly, such a switch can lead to abrupt movement. However, the character remains balanced thanks to the robustness of the control policy. While the \mvae{} model struggles more with motions substantially different from its training dataset, like tracking a dance using the locomotion \mvae{}, it still strives to reproduce the motion as accurately as possible. Figure~\ref{fig:reconstruction_result} shows screenshots of several results. Note that the input motions do not include muscle-related data. The \mvae{} automatically recovers such information by reproducing the motion in simulation. Figure~\ref{Fig:tracking} shows the muscle activation level curves when tracking four different motions.

To demonstrate the capability of our \mvae{} to handle more dynamic and challenging short motions, we train a separate \mvae{} on a \emph{Jump Spin Kick} motion from the SFU Motion Capture Database \cite{sfudataset}. The policy allows the character to perform the skill indefinitely, as sketched in \fig\ref{fig:spin_kick}.
We encourage readers to view the supplemental video for a better visual evaluation of the results. 

\paragraph{Generation} 
We experiment with two generation tasks, both using the trained locomotion \mvae{}. In the \emph{random sampling} task, we draw random latent codes from the state-dependent prior distribution and decode these codes into motion using the policy and simulation. As shown in Figure~\ref{fig:randomwalk}, our \mvae{} generates a diverse range of high-quality motions in this setting. The fluctuating latent codes can cause the character to frequently change its skills, leading to occasional jittering. However, this issue can be mitigated by sampling with smaller noise. 

In the \emph{high-level control} task, we consider a downstream task of controlling the speed and direction of the character. We train a task policy to compute the latent codes with the objective of minimizing the discrepancy between speed and heading direction of the character and input from user. Once trained, this policy allows the character to adjust its direction and speed smoothly in response to interactive user commands. 
Please refer to the supplementary video for a visualization of such behaviors.

\paragraph{Fatigue}
\label{para:fatigue_exp}
In both the tracking and generation tasks, the character adapts its movement based on the evolution of the fatigue stage. Fatigue accumulates naturally based on the muscle activation levels, leading to varied patterns over extended exercises. Figure \ref{Fig:tracking_fatigue} depicts the changes of the fatigue parameter $M_F$ when the character is instructed to track the same motion indefinitely. It can be observed that more dynamic motions can result in faster fatigue accumulation.  

To better demonstrate the fatigue effect, we instruct the character to hold its arms horizontally using the muscle-space PD control. The character's body is fixed to prevent it from falling. Figure \ref{fig:arm_hold_curve} illustrates the changes in activation level and the fatigue parameter over time. As depicted in Figure \ref{fig:arm_hold} and also shown in the supplementary video, the character's arm gradually descends due to fatigue accumulation. When we let the character rest its arm for a certain amount of time (120\~{}180s in \fig\ref{fig:arm_hold_curve}), the fatigued muscle recovers, enabling the character to raise its arm back to its initial height. This experiment can be extended to more complex motions. Specifically, we instruct the character to run for an extended period, causing it to become fatigued and subsequently run in a less powerful manner. After this, the character is directed to walk or stand for a few seconds. During this time, the character's fatigue recovers, allowing it to return to its original running style. Figure~\ref{fig:fatigue_run_test} shows the fatigue curves for two major muscles in both the run-walk and run-stand settings. Notably, muscle fatigue recovers faster when the character is standing than walking, and the character exhibits a better recovery of its motion style in the subsequent running. In this case, we increased fatigue rate by 50 times for faster fatigue manifestation.

We train our \mvae{} using a predefined set of parameters for the fatigue dynamics. However, the trained model can resist changes to such parameters. To demonstrate this, we increase the values of $F$ and $R_r$ in the 3CC-r model to 5, 10 and 25 times that used in training, respectively. The character can still perform locomotion under the control of the trained \mvae{}. However, with the latter settings, it becomes fatigued much faster, causing a more rapid change in motion style.

\subsection{Ablation Study}
{To show the importance of our proposed muscle-space control framework,  we compare its performance with a vanilla muscle control strategy that directly controls the muscles using activation signals. Given that muscle activations are confined to the range of [0, 1], as suggested by \citet{lee2019scalable}, we introduce an additional activation function in the form of $\text{ReLU}(\text{Tanh}(\vect{x}))$ after the last layer of the \mvae{} policy. This modification ensures compliance with the aforementioned range constraint. All other network configurations, including the prior distribution, the posterior distribution, the policy, and the world model, remain unchanged. The physical parameters of the character also remain the same. Figure~\ref{fig:learning_curve_mvae} shows typical learning curves of \mvae{}s with the proposed muscle-space control and the vanilla muscle activation control. The results indicate that the system struggles to find a feasible \mvae{} in the \mvae{} + vanilla activation control setting. The character is unable to maintain balance, leading to an early plateau in the reward without any growth.}

%% file: Sections/conclusions.tex
\section{Conclusion}
\label{sec:conclusion}
In this paper, we present a comprehensive simulation and control framework for muscle-actuated characters. We augment the widely used Hill-type muscle mechanism with the 3CC-r fatigue dynamics model, effectively simulating activation dynamics and fatigue effects. We further propose a muscle-space control mechanism that combines PD control with a simple strategy that equivalently realizes the fatigue dynamics using more efficient clip operations. This control framework allows for the learning of a VAE-based generative control model, the \mvae{}, which can accommodate a diverse range of movements in an unstructured motion dataset. The \mvae{} model enables the encoding of not only motion features but also muscle control and fatigue properties within a rich and flexible latent space. With the aid of the \mvae{}, we can easily recover muscle dynamics information from a motion by tracking it in simulation. Moreover, The muscle-actuated character can generate a variety of motion skills by sampling from the latent space and can accomplish downstream tasks by learning high-level control policies that operate in that latent space. The \mvae{} model incorporates fatigue states into all its components, leading to a natural evolution of motion styles during extended exercises. We believe that our findings extend beyond the realm of graphics and have potential implications in other domains such as biomechanics and human-computer interaction.

Currently, there are several limitations in our framework. First, our simulation model and control strategy encompass numerous parameters. 
Many of these parameters are borrowed from existing literature \cite{lee2019scalable,looft20183ccr,xia2008theoretical} but were originally measured for or designed around human bodies with distinct properties, possibly making them inaccurate for the character we have utilized. A future research avenue could involve automatically determining these parameters from motion data or body measurements. Second, our control model does not account for muscle coordination. All the muscles are currently controlled individually. Despite our results indicating some coordination, there are times when the character activates both agonist and antagonist muscles simultaneously, which is not biomechanically accurate. Incorporating more biomechanical aspects, like relationships among agonists, antagonists, and synergists during a movement, could increase the control's biomechanical precision. Third, \mvae{} aims to replicate reference motions faithfully, but if these motions are not physically viable, the resultant motion can produce unsatisfactory artifacts. For instance, because of retargeting errors, the character's legs sometimes intersect in our training motions. If we track these motions with self-collision activated, the character may trip and fall. While fine-tuning the policy with self-collision helps the character maintain balance post-trip, it does not rectify the issue present in the reference motion. Another example is our training of \mvae{} on a manually crafted \emph{Horse Stance} motion. 
Its imbalance causes a conflict between pose tracking and balance maintenance, making the character oscillate and result in a waggling motion. Lastly, this training objective encourages the character to reproduce training motions without considering its fatigue level. This often leads the character to sustain its motion until it loses balance. Fatigue should not only be perceived as a mechanical effect but also as a stylistic element to motion. Exploring the transition from mechanical changes due to fatigue to the consequent changes in style is a noteworthy research challenge.

%% file: Sections/acks.tex
\begin{acks}
We would like to thank Baoquan Chen for his insightful discussions and assistance. We are also thankful to Zhenhua Song for his invaluable suggestions on the rendering system and rigid body simulation, and to Heyuan Yao for his inspiring conversations on generative models and his organization of open-source code. We also appreciate the anonymous reviewers for their constructive feedback. This work was supported in part by The Fundamental Research Funds for the Central Universities, Peking University.
\end{acks}

%% file: Sections/x_figure_only_page.tex
\clearpage

\begin{figure*}[h]   

    \begin{subfigure}{0.496\linewidth}
      \includegraphics[width=1.005\linewidth]{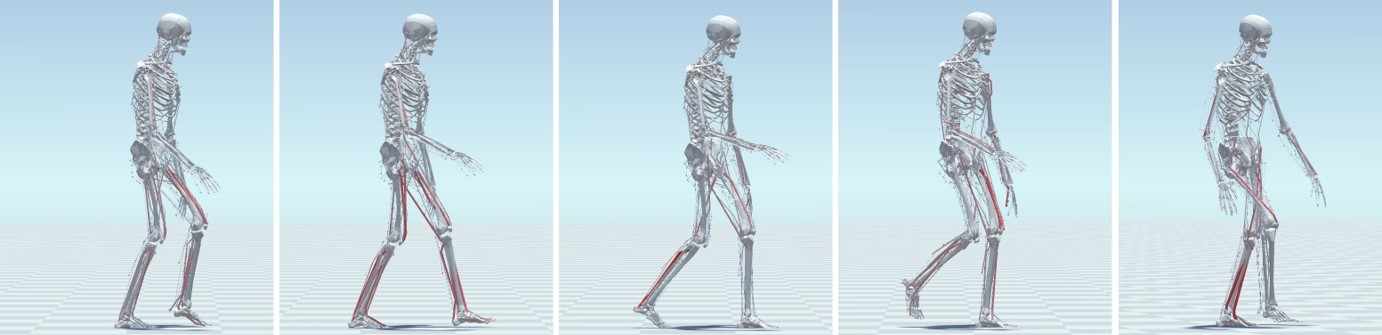}
      \caption{Walk}
    \end{subfigure}
    \begin{subfigure}{0.496\linewidth}
      \includegraphics[width=1.013\linewidth]{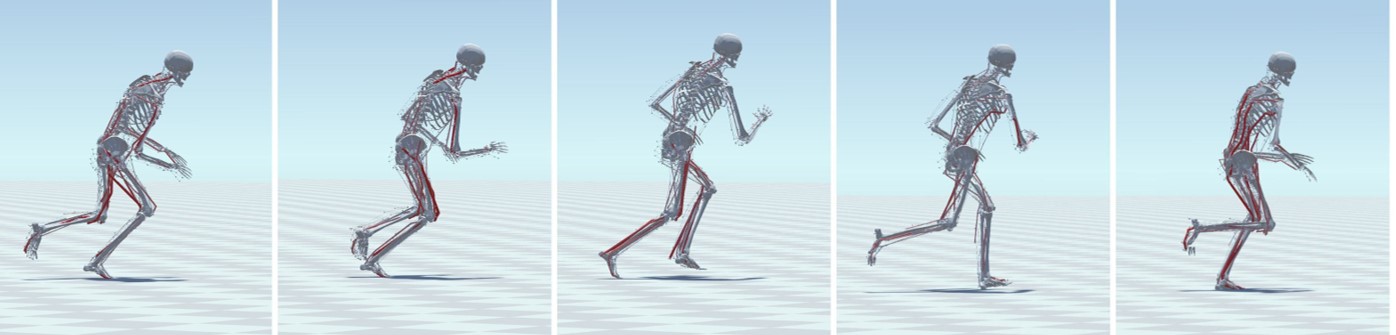}
      \caption{Run}
    \end{subfigure}
  
    \begin{subfigure}{\linewidth}
      \includegraphics[width=\linewidth]{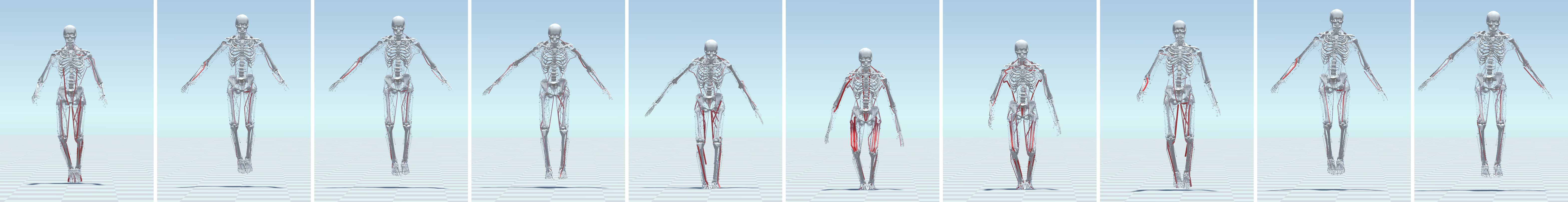}
      \caption{Jump}
    \end{subfigure}
    \begin{subfigure}{\linewidth}
      \includegraphics[width=\linewidth]{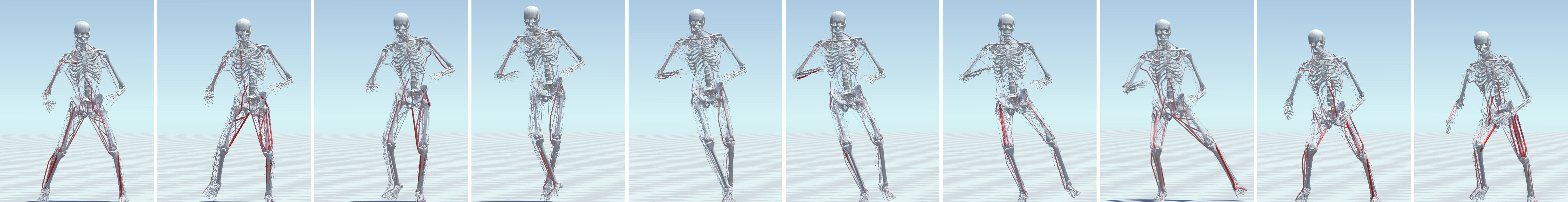}
      \caption{Side Walk}
    \end{subfigure}
    
    \begin{subfigure}{\linewidth}
      \includegraphics[width=\linewidth]{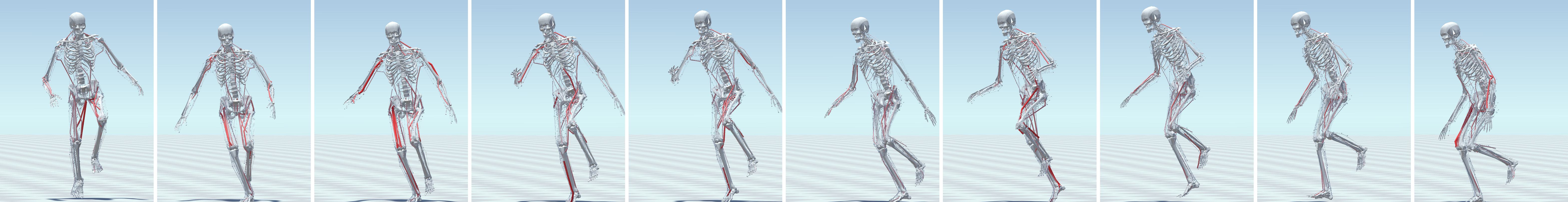}
      \caption{Single Jump}
    \end{subfigure}
    \begin{subfigure}{\linewidth}
      \includegraphics[width=\linewidth]{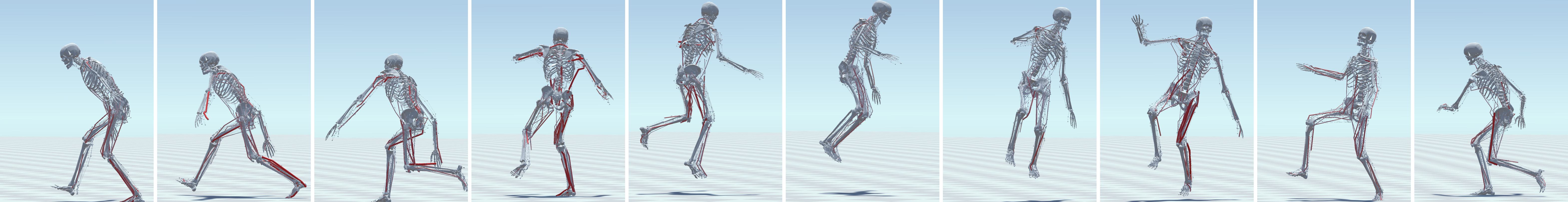}
      \caption{Spin Jump}
    \end{subfigure}
    \begin{subfigure}{\linewidth}
      \includegraphics[width=\linewidth]{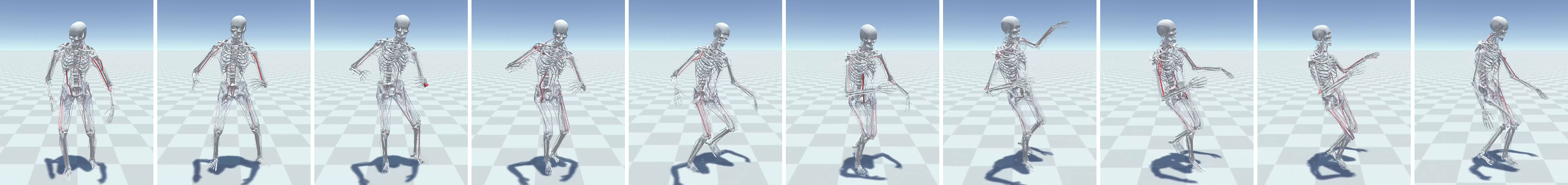}
      \caption{Dance}
      \label{fig:track_dance}
    \end{subfigure}
    \begin{subfigure}{\linewidth}
      \includegraphics[width=\linewidth]{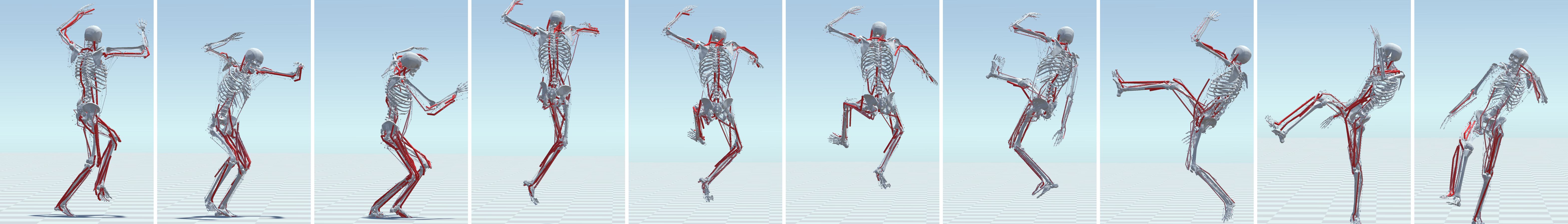}
      \caption{Jump Spin Kick}
      \label{fig:spin_kick}
    \end{subfigure}    
    \Description{}
    \caption{The muscle-actuated character can perform a diverse range of motions.}
    \label{fig:reconstruction_result}
\end{figure*}

\begin{figure}[t]
  \centering
  \includegraphics[width=0.95\linewidth]{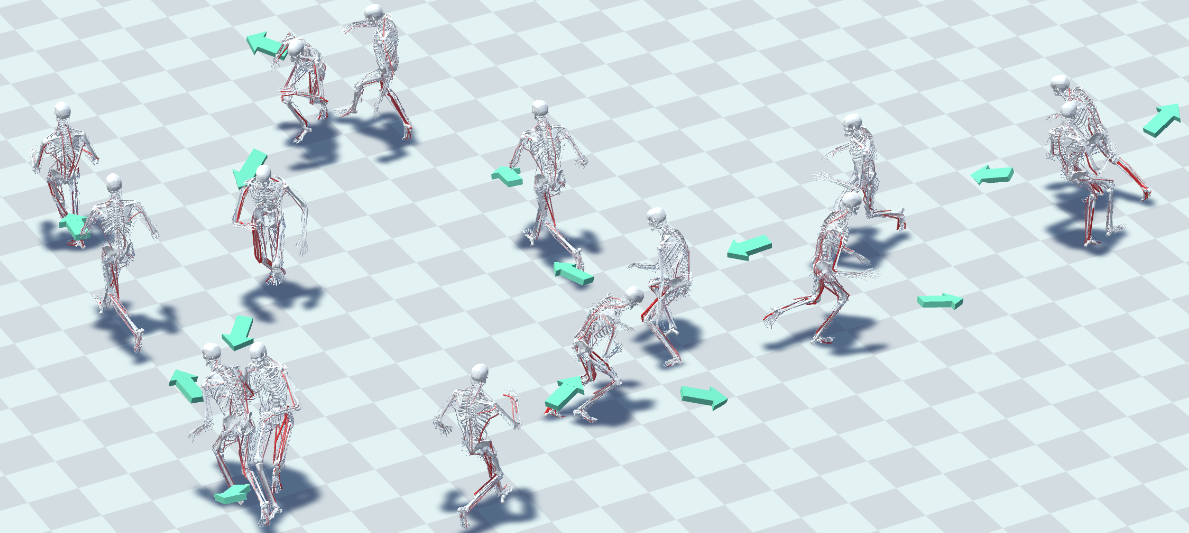}
  \caption{Trajectories generated by the muscle-actuated character in the random sampling experiment. The arrows indicate moving directions.}  
  \Description{}
  \label{fig:randomwalk}
\end{figure}
\begin{figure}[t]
  \centering
  \includegraphics[width=1.0\linewidth]{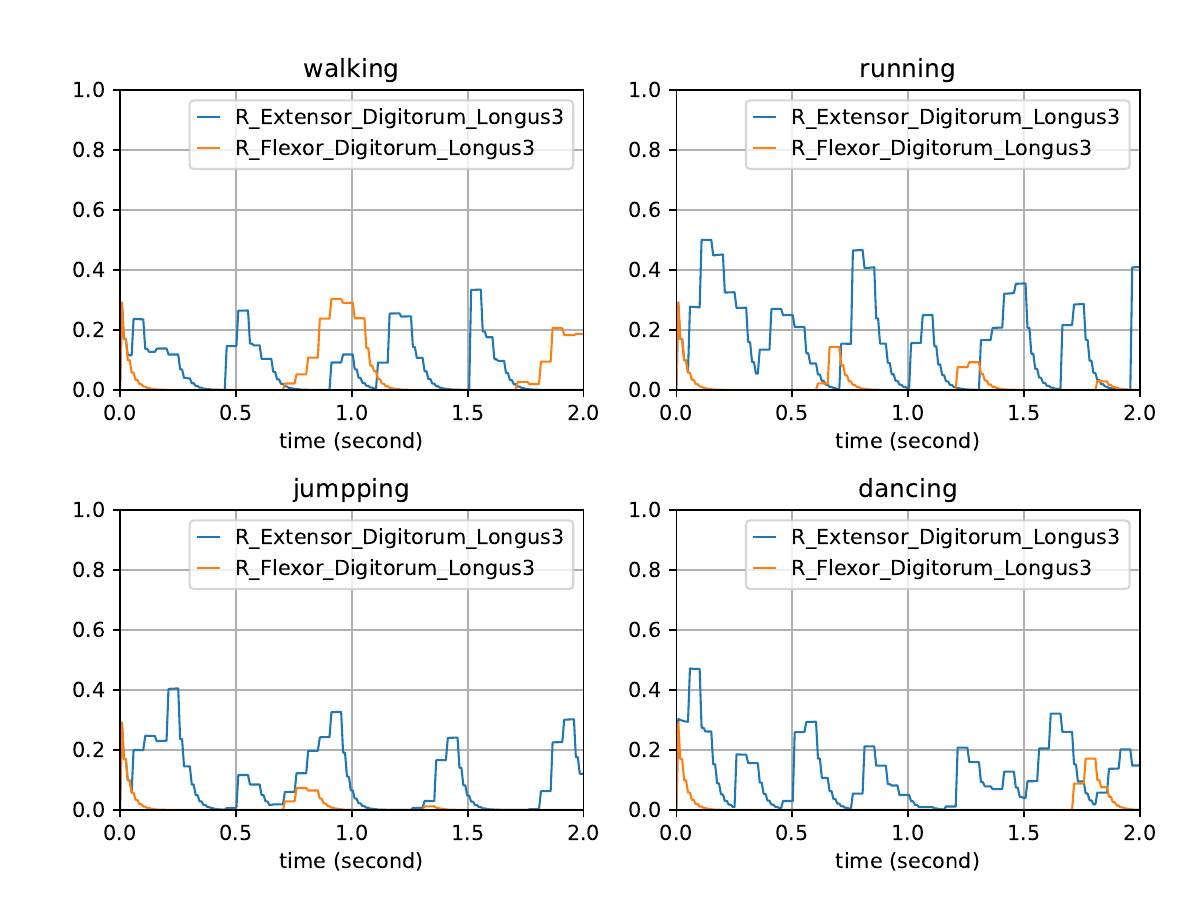}
  \caption{Muscle activation curves of different motions.}
  \Description{}
  \label{Fig:tracking}
\end{figure}

\begin{figure}[t]
  \centering
  \includegraphics[width=1.0\linewidth]{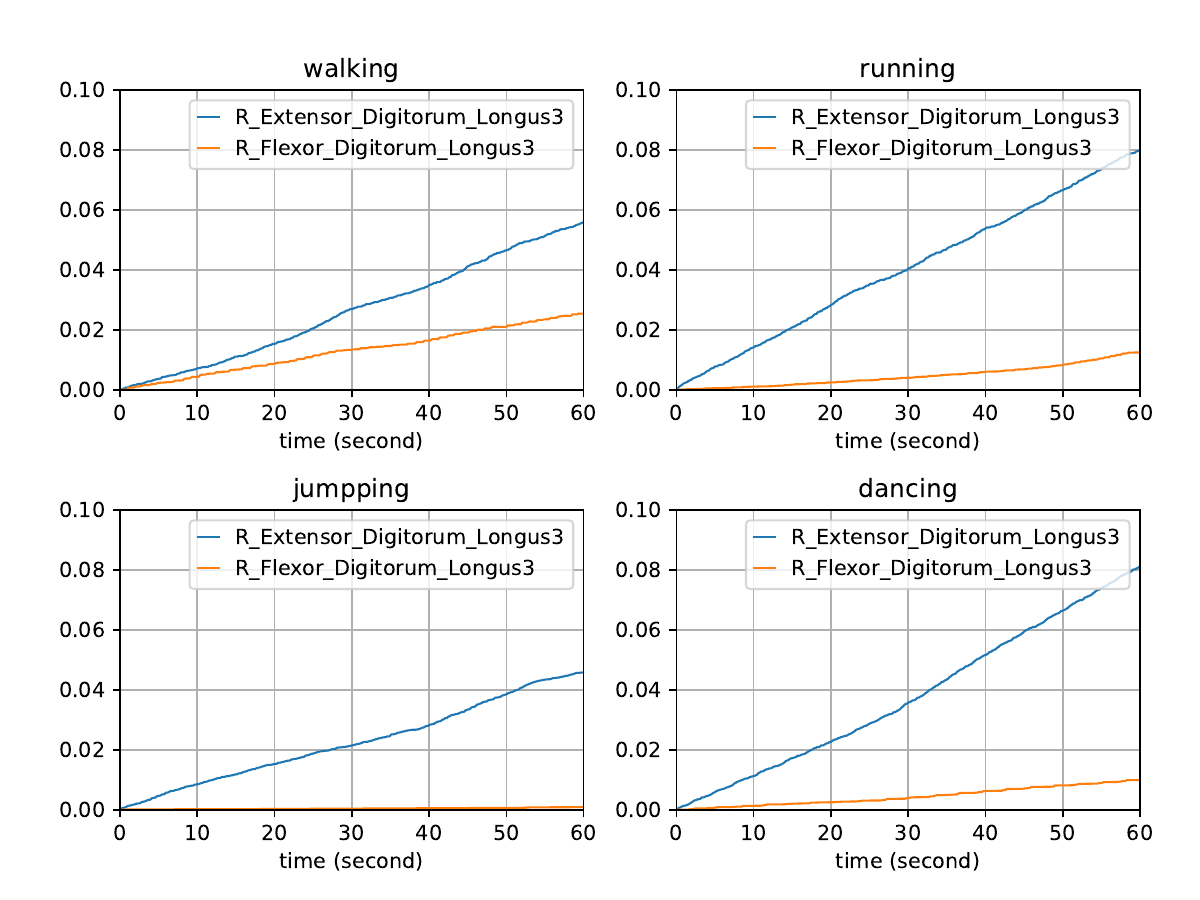}
  \caption{Muscle fatigue curves of different motions.}
  \Description{}
  \label{Fig:tracking_fatigue}
\end{figure}

\begin{figure}[t]
  \centering
  \includegraphics[width=\linewidth]{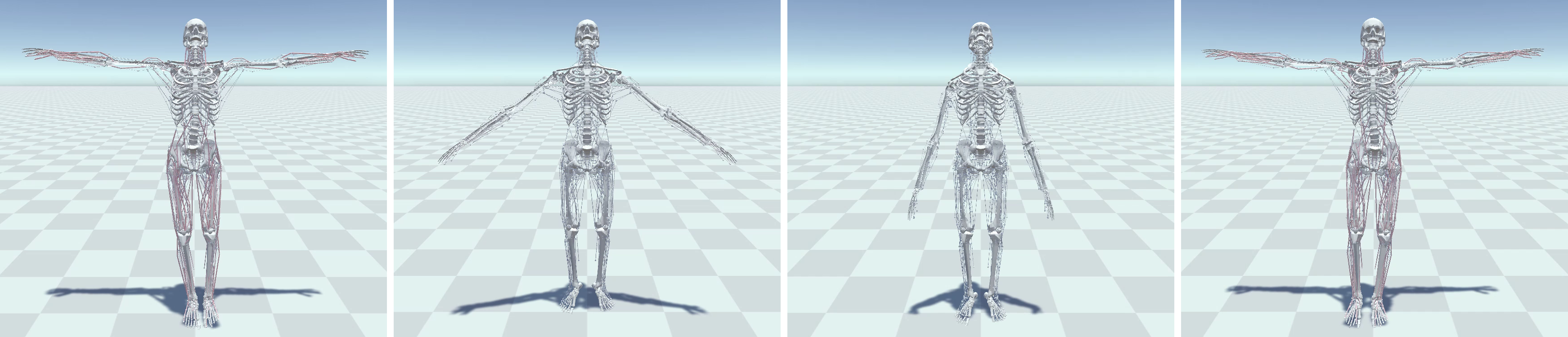}
  \caption{The character holds its arms horizontally but fatigues after 2 minutes. After a 1-minute rest, it resumes the task, recovering from fatigue. }
  \Description{}
  \label{fig:arm_hold}
\end{figure}

\begin{figure}[t]
  \centering
  \includegraphics[width=0.8\linewidth]{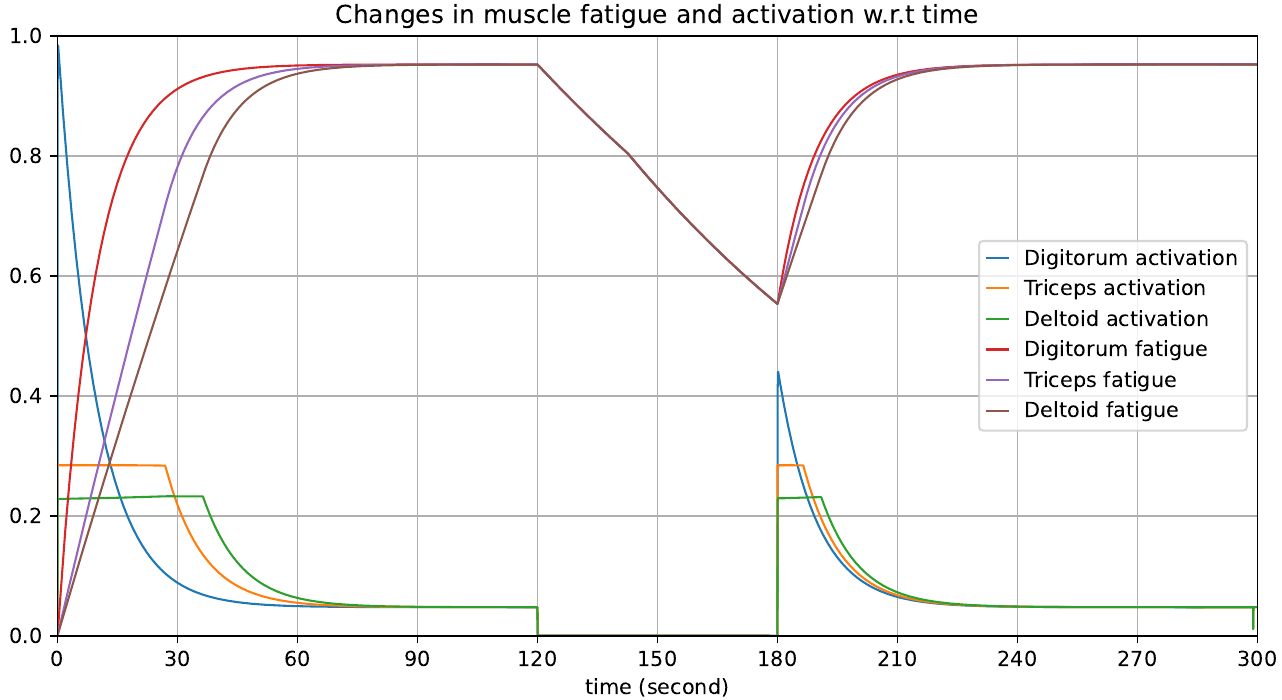}
  \caption{Activation and fatigue curves of the forearm, upper arm, and shoulder muscles in the experiments shown in \fig\ref{fig:arm_hold}.}  
  \Description{}
  \label{fig:arm_hold_curve}
\end{figure}

\begin{figure}
  \includegraphics[width=0.49\linewidth]{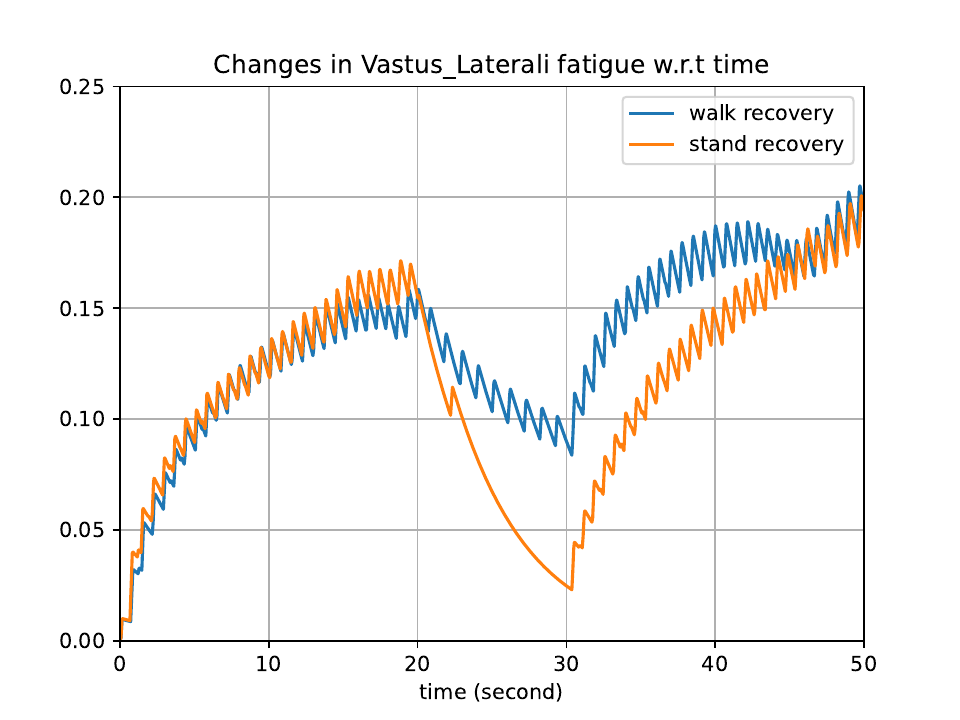}
  \includegraphics[width=0.49\linewidth]{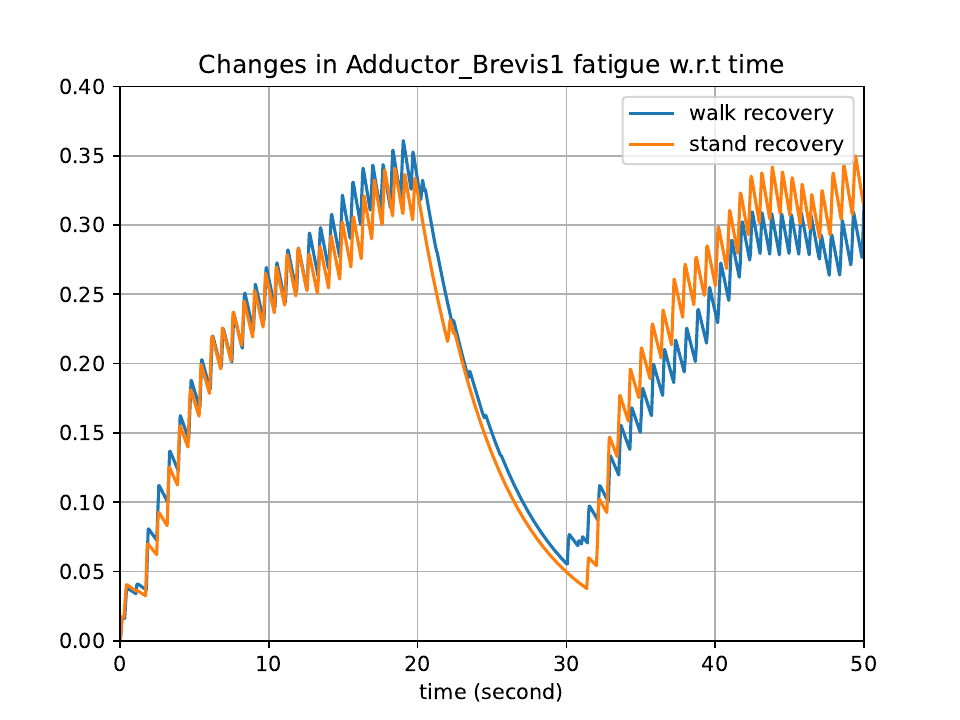}
  \caption{Fatigue curve of the fatigue and recovery in the run-walk/idle test.}
  \label{fig:fatigue_run_test}
\end{figure}

\begin{figure}[t]
  \centering 
  \includegraphics[width=0.9\linewidth]{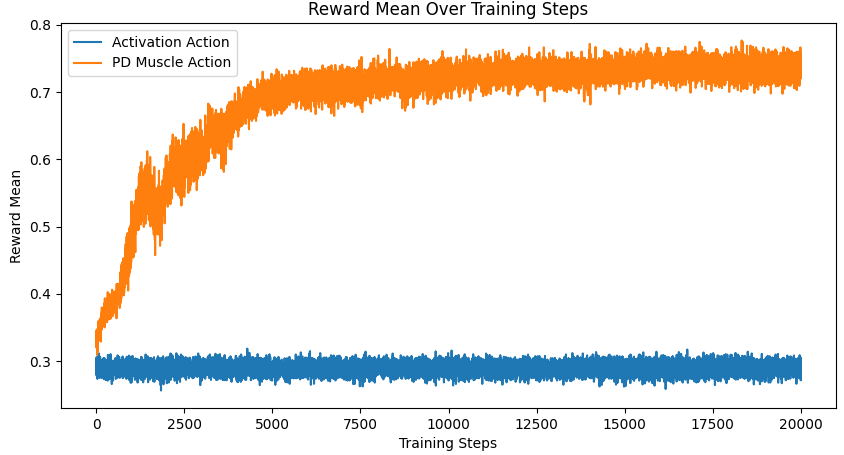}
  \caption{Typical learning curves of \mvae{} using the proposed muscle-space control and the vanilla muscle activation control. }
  \label{fig:learning_curve_mvae}
\end{figure}

\clearpage

%% file: Sections/supplimentary.tex
\appendix
\clearpage

\section{Muscle Modeling}
\subsection{Muscle Routing}
\label{muscle_routing_appendix}
We approximated the muscles of the Hill-type model as polylines, with the inflection points of these polylines serving as the muscles' anchor points. The positioning of the muscle anchors can be characterized by LBS (Linear Blend Skinning), as mentioned in the main article. These anchor points play a crucial role in determining the muscle's route and also provide the location where the mid-muscle contraction force can be transferred to the bones. Specifically, the position of the anchor point is
\begin{equation}
    \vect{p}= \sum w_iT_i\vect{x}_i' ,
    \label{equ:lbs}
\end{equation}
where $\vect{p}$ denotes the position of the muscle anchor point, $\vect{x}_i'$ represents the relative position of the anchor point to the $i$-th bone. The local coordinate system of the $i$-th bone can be represented as a translation-rotation matrix, denoted as $T_i$, in the global coordinate system. The variable $w_i$ is the weight of the anchor point with respect to the i-th bone, which is computed based on the distance between the anchor point and the bone, as suggested by \citet{jehee19muscle}. Leveraging the anchor positions, we can calculate the muscle length as
\begin{equation}
l_{\mathrm{M}}=\sum_{k=1}^{n-1}\left\|\vect{s}_k\right\| , \ \vect{s}_k=\vect{p}_{k+1}-\vect{p}_k ,
\label{equ:muscle_length}
\end{equation}
where $n$ is the number of anchor points of the muscle. 

We utilize a muscle dynamics model to compute the amplitude of a muscle's force, denoted by $f_m$, as detailed in the next section. The forces applied at each anchor point are then computed as
\begin{equation}
    \forcerm_k^{-}=f_{\mathrm{m}}\frac{\point_{k-1}-\point_k}{\left\|\point_{k-1}-\point_k\right\|}, \quad \forcerm_k^{+}=f_{\mathrm{m}} \frac{\point_{k+1}-\point_k}{\left\|\point_{k+1}-\point_k\right\|} .
\end{equation}
In this expression, $\vect{f}_k^{-}$ and $\vect{f}_k^{+}$ represent the forces exerted along the two polylines that join at the $k$-th anchor point of the muscle, respectively. Notably, for endpoint anchors, both $\vect{f}_0^{-}$ and $\vect{f}_n^{+}$ are set to zero. All these forces are applied to the bones, which are simulated as rigid bodies, at the corresponding anchor points to drive the character's motion.

\subsection{Muscle Dynamics}

\begin{figure}
  \centering
  \includegraphics[width=\linewidth]{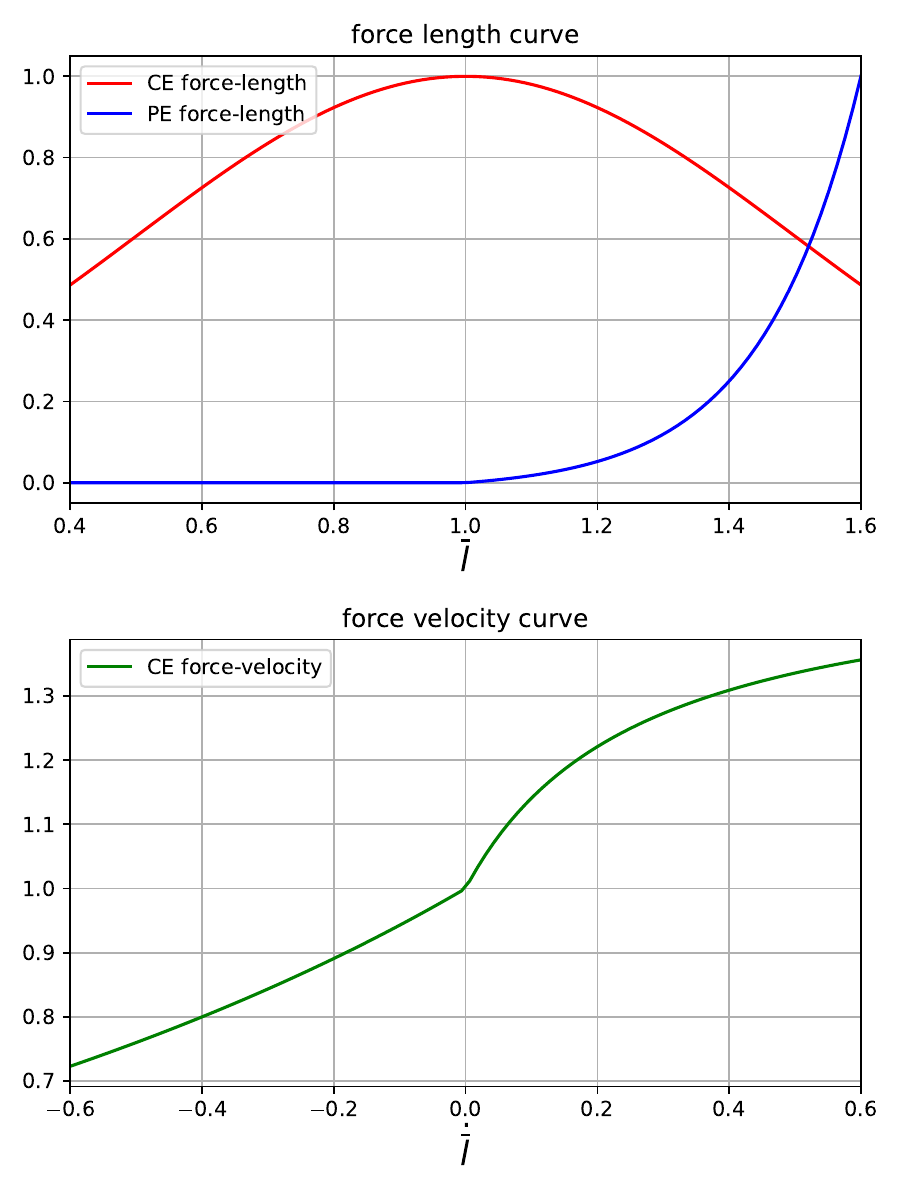}
  \caption{The force-length and force-velocity curves used in our experiment. Curves of the active muscle force are drawn in red and those of the passive muscle force are in blue. 
  }
  \Description{}
  \label{fig:muscle_hill_type}
\end{figure}

The Hill-type muscle model \cite{hill1938heat, zajac1989muscle} is a widely adopted approach in character animation \cite{geijtenbeek2013flexible, jehee19muscle}. This model comprises a contractile element (CE), a parallel elastic element (PE), and a tendon element. We simplify this model by neglecting changes in tendon length and the pennation angle following the work which are widely accepted in computer animation \cite{jiang2019synthesis, jehee19muscle, geijtenbeek2013flexible}. In the main article, we compute contractile muscle force as
\begin{equation}
  f_{\text{m}} = \atv f_{\text{CE}}^l(\bar{l})f_{\text{CE}}^v(\dot{\bar{l}})+f_{\text{PE}}(\bar{l}),
\label{equ:hill-1}
\end{equation}
which is a compact version of the formulation 
\begin{equation}
    f_{\text{m}} = f_{\text{m0}}\left(\atv \bar{f}_{\text{CE}}^l(\bar{l})\bar{f}_{\text{CE}}^v(\dot{\bar{l}})+\bar{f}_{\text{PE}}(\bar{l})\right), 
    \label{equ:hill_supp}
\end{equation}
Here $f_{\text{m0}}$ is the maximum isometric force, which is determined by the type, size, and several other properties of a muscle. The force-length and force-velocity functions, i.e., $\bar{f}_{\text{CE}}^l(\bar{l})$, $\bar{f}_{\text{CE}}^v(\dot{\bar{l}})$, and $\bar{f}_{\text{PE}}(\bar{l})$, are assumed to be the same for all the muscles. The normalized muscle length and its rate of change, $\bar{l}$ and $\dot{\bar{l}}$, are computed as
\begin{align}
\bar{l} &= \frac{ {l_{\text{M}}}/{l_{\text{ori}}} - l_{\text{Tnorm}} }{l_{\text{MTnorm}}} \\
\dot{\bar{l}} &= \frac{1}{\dt}\left( \bar{l}^t - \bar{l}^{t-1} \right) ,
\label{equ:normalized_len}
\end{align}
where $l_{\text{ori}}$ is the rest length of the muscle. $l_{\text{Tnorm}}$ and $l_{\text{MTnorm}}$ are the normalizing factors of tendon length and muscle-tendon unit length of the muscle, respectively. The values of $f_{\text{m0}}$ and these normalizing factors for each muscle can be found in biomechanics literature, such as \cite{delp2007opensim}. In this paper, we borrow these values from \cite{lee2019scalable}. The formulation of the functions $\bar{f}_{\text{CE}}^l$, $\bar{f}_{\text{CE}}^v$, and $\bar{f}_{\text{PE}}$ used in this paper are:  
\begin{equation}
\begin{aligned}
  \bar{f}_{\text{CE}}^l(\bar{l}) &= \exp \left( -\dfrac{(\bar{l} - 1)^2}{0.5} \right) \\
  \bar{f}_{\text{CE}}^v(\dot{\bar{l}}) &=\begin{cases} 
    1.5 + \dfrac{0.5\times (-10.0 + \dot{\bar{l}})}{37.8\dot{\bar{l}} + 10.0} & \text{if } \dot{\bar{l}} > 0.0   \\
    \dfrac{-10 - \dot{\bar{l}}}{-10.0 + 5.0\dot{\bar{l}}} & \text{otherwise } 
  \end{cases}\\
  \bar{f}_{\text{PE}}(\bar{l}) &= \begin{cases} 
    \frac{\exp\left(\dfrac{4.0\times (\bar{l} - 1.0)}{0.6}\right) - 1.0}{\exp(4.0) - 1.0} & \text{if } \bar{l} > 1.0 \\
    0.0 & \text{otherwise}
  \end{cases}
\end{aligned}.
\label{equ:muscle_func}
\end{equation}
Figure \ref{fig:muscle_hill_type} shows the graphs of these functions.

\subsection{Fatigue Dynamics}

\begin{figure}
  \includegraphics[width=0.7\linewidth]{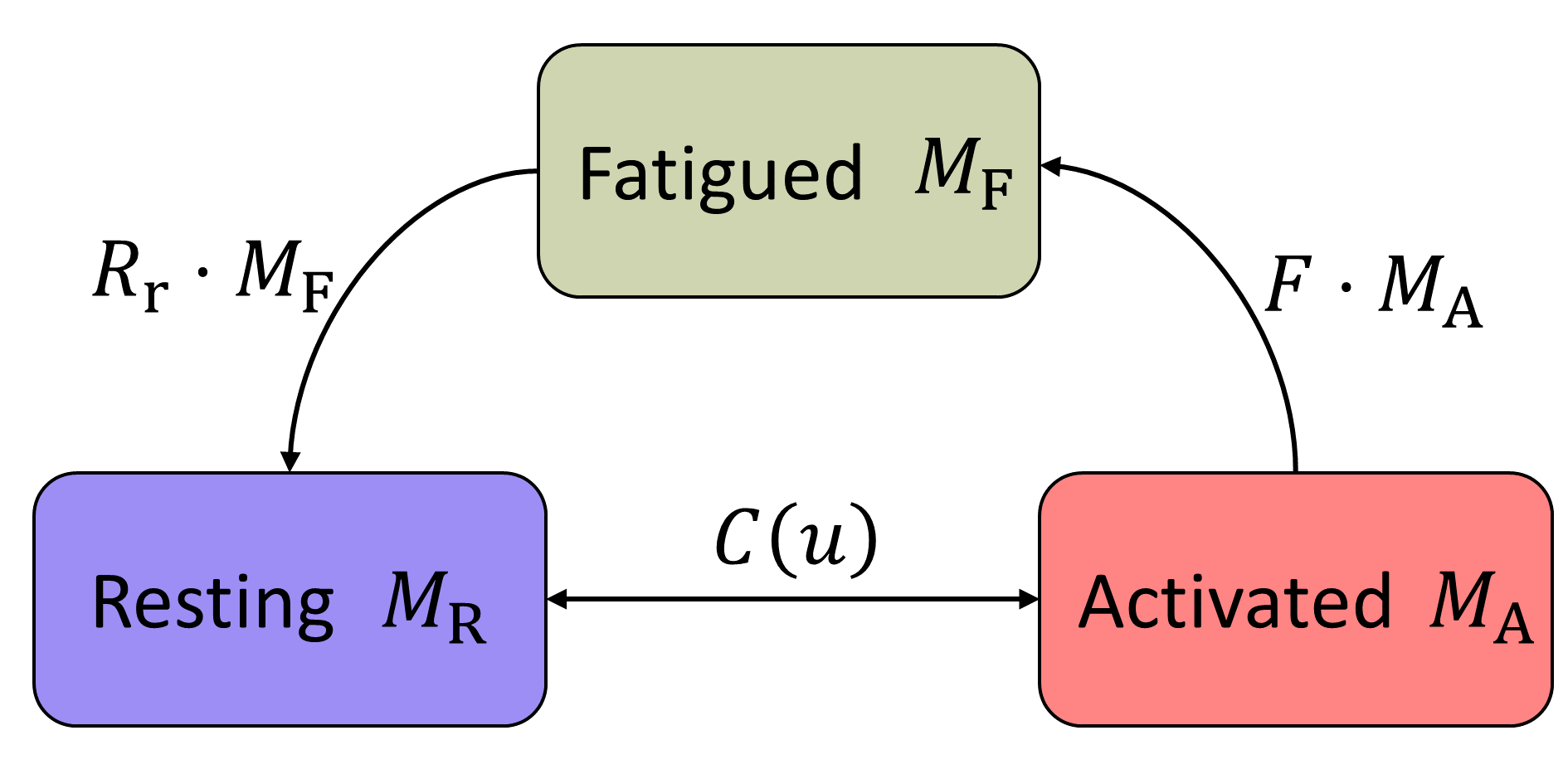}
  \caption{3CC-r assumes muscle actuators to be in one of three possible states. These states are governed by a set of differential equations.} 
  \label{fig:3ccr_model}
  \Description{}
\end{figure}

We adopt 3CC-r model \cite{looft20183ccr} as the fatigue dynamics model, which is an enhanced version of the Three Compartment Controller (3CC) model proposed by \citet{xia2008theoretical}. 
The 3CC-r model assumes that each muscle consists of multiple hypothetical muscle-tendon actuators. Each of these actuators is presumed to be in one of three possible states (compartments):
\begin{itemize}
    \item \textbf{Activated} $M_\text{A}$: The muscle actuator is contributing.
    \item \textbf{Resting} $M_\text{R}$: The muscle actuator is inactivated but can be recruited.
    \item \textbf{Fatigued} $M_\text{F}$: The muscle actuator is fatigued and cannot be utilized.
\end{itemize}
We employ a unit-less measure of muscle force, expressed as a percentage of the maximum voluntary contraction (MVC), to describe the effect of fatigue, following existing literature. The values $M_\text{A}$, $M_\text{R}$, and $M_\text{F}$ are expressed as percentages of MVC. The resting muscle actuator ($M_\text{R}$) is recruited to become an activated muscle actuator ($M_\text{A}$) when there is a load requirement. Once activated, the muscle actuator's power decays and fatigue accumulates. The transition relationships among the three states of the muscle are illustrated in Figure \eqref{fig:3ccr_model}. The following equations describes the change of these values over time for each compartment:
\begin{equation}
\begin{aligned}  
& \frac{\mathrm{d} M_\text{A}}{\mathrm{d} t}=C(\targetload)-F \cdot M_\text{A} \\
& \frac{\mathrm{d} M_\text{R}}{\mathrm{d} t}=-C(\targetload)+R_{\text{r}} \cdot M_\text{F} \\
& \frac{\mathrm{d} M_\text{F}}{\mathrm{d} t}=F \cdot M_\text{A}-R_{\text{r}} \cdot M_\text{F}
\end{aligned}
\label{equ:3cc_supp}
\end{equation}
where $F$ and $R_r$ denote the fatigue and recovery coefficients. $r$ is an additional rest recovery multiplier introduced by \citet{looft20183ccr}, which alters the recovery coefficient as
\begin{equation}
    R_{\text{r}}= \begin{cases}
      r \cdot R &  M_\text{A} \geq \targetload \\ 
      R &  M_\text{A} <  \targetload 
    \end{cases}
\end{equation}
The function $C(\targetload)$ in \eqn \eqref{equ:3cc_supp} dynamically change the ratio of $M_\text{A}$ and $M_\text{R}$ based on the target load $\targetload$. It is formulated as a piecewise linear function that increases monotonically with $\targetload$:
\begin{equation}
  C(u)= \begin{cases}
    L_{\text{R}} \cdot\left(\targetload-M_\text{A}\right) & \text { if } M_\text{A} \geq \targetload \\ 
    L_{\text{D}} \cdot\left(\targetload-M_\text{A}\right) & \text { if } M_\text{A}<\targetload \text { and } M_\text{R}>\targetload-M_\text{A} \\ 
    L_{\text{D}} \cdot M_\text{R} & \text { if } M_\text{A}<\targetload \text { and } M_\text{R} \leq \targetload-M_\text{A} ,
  \end{cases}
  \label{equ:ct_in_3cc}
\end{equation}
which is characterized by the development factor $L_{\text{D}}$ and relaxation factor $L_{\text{R}}$. It is worth noting that $C(\targetload)$ also depends on the current activation level $M_\text{A}$. This relationship effectively prevents the activation level from changing instantaneously, thereby replicating the behavior of activation dynamics~\cite{Thelen2003Adjustment,winters1995}.

The target load, $\targetload$, represents the effort the brain expects the musculoskeletal system to generate, resulting from the combined effects of physiological and neurological processes. \targetload{} can also be depicted as a normalized, unit-less coefficient that reflects the percentage of actuators a muscle is required to recruit, thus $\targetload\in[0,1]$. Furthermore, \targetload{} effectively functions as the muscle excitation in the activation dynamics model~\cite{Thelen2003Adjustment,winters1995}. Notably, the target load \targetload{} is allowed to change instantaneously within the range of $[0, 1]$. However, due to the existence of muscle fatigue, it may not always be realizable by the musculoskeletal system.

\begin{figure}
  \includegraphics[width=0.98\linewidth]{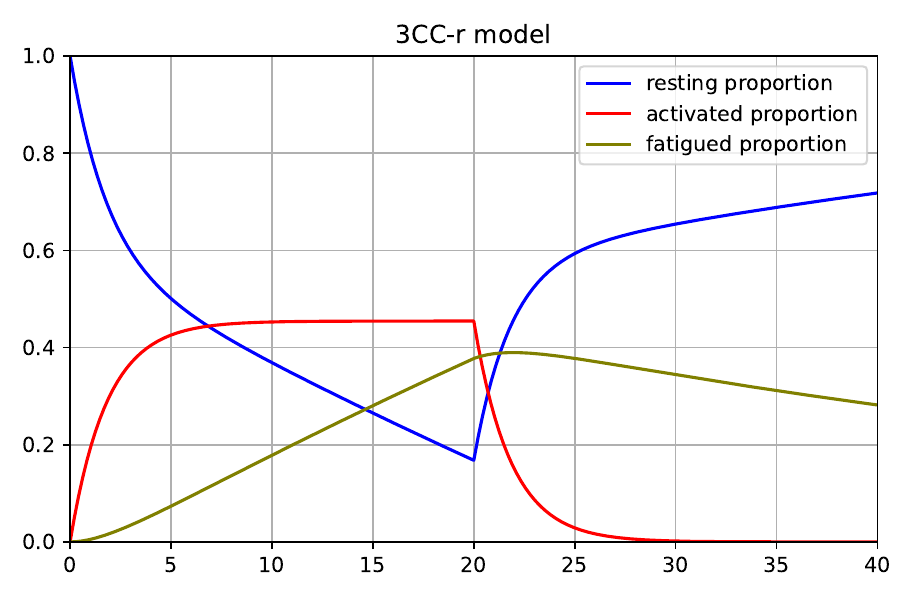}
  \caption{An example of 3CC-r to illustrate the evolution of $M_{\text{A}}$ (red), $M_{\text{R}}$ (blue), $M_{\text{F}}$ (chartreuse) under the square wave target load stimuli. The muscle is set at non-fatigued and non-activated state at the beginning.}
  \label{fig:3ccr_demo}
  \Description{}
\end{figure}

We provide an example in Figure \ref{fig:3ccr_demo} to illustrate the evolution of the components in the 3CC-r model. Initially, the muscle is in a non-fatigued and non-activated state with $M_{\text{A}}=M_{\text{F}}=0.0$ and $M_{\text{R}}=1.0$. The target load \targetload{} is set to 0.5 for the first 20 seconds and then reset to 0. Note that Figure \ref{fig:3ccr_demo} merely serves as an illustration, the values of $L_R$, $L_D$, $F$, and $R_r$ are adjusted to enhance the visibility of the curves.

\subsubsection{Fatigue dynamics as clip operation}
Our muscle-space PD control calculates a desired force for each muscle. However, due to muscle dynamics and fatigue, these forces are not always achievable. As discussed in the main article, our strategy is to clip these desired forces to within feasible ranges and then apply the resulting forces to the character. Here, we derive the equations used to determine these feasible ranges.

The Hill-type muscle model given in \eqn\eqref{equ:hill-1} suggests that the muscle force monotonically increases with respect to the activation level. We can then compute the desired muscle activation $\atv_{\text{pd}}$ based on the desired force $\force_{\text{pd}}$ using
\begin{equation}
  \atv_{\text{pd}} = \frac{\force_{\text{pd}} - f_{\text{PE}}(\bar{l})}{f_{\text{CE}}^l(\bar{l})f_{\text{CE}}^v(\dot{\bar{l}})} \quad.
  \label{equ:alpha_pd_supp}
\end{equation}
Notably, $\atv_{\text{pd}}$ may not be achievable due to the muscle constraints and fatigue.

As discussed in the main article, $M_{\eqword{A}}$ and $\atv$ are equivalent because they both represent the muscle activation level. The first equation in \eqn~\eqref{equ:3cc_supp} can be rewritten as
\begin{equation}
  \dot{\atv}=C(\targetload)-F\atv ,
\end{equation}
which can be discretized using the forward Euler method. Denoting the muscle activation in the subsequent time step as $\tilde{\atv}$, we have
\begin{equation}
  \dot{\atv} \approx \frac{\tilde{\atv}-\atv}{\dt} , \quad \text{or}, \quad
  \tilde{\atv} \approx \dot{\atv}\dt + \atv .
  \label{equ:activation_dot}
\end{equation}
\begin{equation}
  \text{So,} \quad
  \tilde{\atv}=\tilde{\atv}(\targetload)=K\atv + \dt{}C(\targetload) ,
  \label{equ:atv_u_supp}
\end{equation}
where $K=1-\dt{}F$ can be considered as a decay factor. With \eqn\eqref{equ:atv_u_supp}, our objective is now to find a target load $\targetload$ within its feasible range, $[0,1]$, that can leads to a feasible $\tilde{\atv}$ close to $\atv_{\text{pd}}$.

Substituting \eqn\eqref{equ:ct_in_3cc} into \eqn\eqref{equ:atv_u_supp}, we get
\begin{equation}
\tilde{\atv}(\targetload)= \begin{cases}
  K\atv + \dt L_{\text{R}} \left(u-\atv\right) & \targetload\le\atv \\ 
  K\atv + \dt L_{\text{D}} \left(u-\atv\right) & \atv < \targetload<\alpha+M_{\eqword{R}} \\ 
  K\atv + \dt L_{\text{D}} M_{\eqword{R}} & \targetload \ge \atv+M_{\eqword{R}}.
  \end{cases}
\label{equ:at_1_in_3cc_reformulate}
\end{equation}
It is easy to verify that $\tilde{\atv}(\targetload)$ is continuous and monotonically non-decreasing with respect to $\targetload$. The minimum and maximum values of $\tilde{\atv}(\targetload)$, given the current activation level $\atv$, are:
\begin{align}
  \tilde{\atv}_{\eqword{lb}}=\tilde{\atv}(0) &= \max(0, K\atv - \dt L_{\text{R}} \atv )\\
  \tilde{\atv}_{\eqword{ub}}=\tilde{\atv}(1) &= \min(1, K\atv + \dt L_{\text{D}} M_{\eqword{R}}) .
\end{align}
Here we use the facts that $\tilde{\atv},\atv\in[0,1]$ and $\atv+M_{\eqword{R}}\in[0,1]$.
Now, we can calculate the feasible $\tilde{\atv}$ that is close to $\atv_{\text{pd}}$ using the clip operator
\begin{equation}
  \tilde{\atv}^* = \text{clip}(\atv_{\text{pd}}, \tilde{\atv}_{\eqword{lb}}, \tilde{\atv}_{\eqword{ub}}) = \begin{cases}
    \tilde{\atv}_{\eqword{lb}} &  \atv_{\text{pd}}\le\tilde{\atv}_{\eqword{lb}} \\ 
    \atv_{\text{pd}}    &  \tilde{\atv}_{\eqword{lb}} < \atv_{\text{pd}} <  \tilde{\atv}_{\eqword{ub}} \\ 
    \tilde{\atv}_{\eqword{ub}} &  \atv_{\text{pd}}\le\tilde{\atv}_{\eqword{ub}}.
    \end{cases}
\end{equation}
Equivalently, we can clip the PD muscle force directly as described in Section~\ref{sec:muscle_actor}. 

After finding the feasible activation level $\tilde{\alpha}^*$, we can further calculate the corresponding $\targetload^*$ that leads to it and use $\targetload^*$ to simulate the 3CC-r model. However, considering that the governing equations in \eqn\eqref{equ:3cc_supp} only depend on $C(\targetload)$, we do not need to explicitly compute $\targetload^*$ but can compute $C(\targetload^*)$ by inverting \eqn\eqref{equ:atv_u_supp}. Specifically,
\begin{equation}
  C(\targetload^*) = \frac{\tilde{\atv}^* - K\atv}{\dt},
\end{equation}
which is used to update $M_\text{R}$ in \eqn\eqref{equ:3cc_supp} using the forward Euler method. In the meanwhile, $M_\text{F}$ in \eqn\eqref{equ:3cc_supp} is updated using the the current muscle activation $\atv$ and $M_\text{F}$.

\section{Muscle VAE} 
 
\subsection{Neural Network Structure}
We formulate the components of the MuscleVAE, specifically the policy $\policy(\act|\stt,\latent)$, the posterior distribution $q(\latent|\stt,\tilde{\stt}_{\text{skeleton}})$, and the state-dependent prior distribution $p(\latent|\stt)$, as normal distributions in the form of $\mathcal{N}(\vect{\mu}_*(~\bigcdot~;\theta_*),\sigma_*^2\vect{I})$. Here, $\sigma_*$ is a predefined standard deviation, and the mean $\vect{\mu}_*(~\bigcdot~;\theta_*)$ is represented by a neural network with trainable parameters $\theta_*$. 
We utilize a latent space $\Latent$ with a dimension of $64$ to encode both motion skills and fatigue style. 

The state-conditional prior distribution is formulated as 
\begin{align}
    p(\latent|\stt) \sim \mathcal{N}\left(\vect{\mu}_p(\stt;{\theta}_p), \sigma_p^2\vect{I}\right),
    \label{eqn:prior_dist}
\end{align}
where $\sigma_{p}=0.3$, $\vect{\mu}_p$ is a neural network with parameters ${\theta}_p$.
The posterior distribution $q(\latent|\stt,\tilde{\stt}_{\text{skeleton}})$  is also a normal distribution 
\begin{align}
    q(\latent|\stt, \tilde{\stt}_{\text{skeleton}}) \sim \mathcal{N}\left(\hat{\vect{\mu}}_q, \sigma_q^2\vect{I}\right)
    \label{eqn:post_dist}
\end{align}
We ensure $q(\latent|\stt, \tilde{\stt}_{\text{skeleton}})$ to be close to the prior $p(\latent|\stt)$ with the same standard deviation $\sigma_{q}=\sigma_{p}=0.3$ and, following the technique used by ControlVAE~\cite{controlvae2022}, formulate the mean of the posterior distribution using a trainable offset function:
\begin{align}
    \hat{\vect{\mu}}_q=\vect{\mu}_p(\stt) + \vect{\mu}_q(\stt, \tilde{\stt}_{\text{skeleton}};{\theta}_q)
\end{align}
where $\theta_q$ represent a collection of neural network parameters. Notebly, with this formulation, the KL-divergence loss in \eqn\eqref{eqn:kl_loss} of the main article has a simpler form:
\begin{align}
    \loss_{\eqword{kl}} &= \sum_{t=0} \gamma^t {\left\Vert\vect{\mu}_q({\stt}^t, \tilde{\stt}^{t+1}_{\text{skeleton}}) \right\Vert_2^2 }/{2{\sigma}_p^2}
    \label{eqn:kl_loss_supp}.
\end{align}
Both $\vect{\mu}_p$ and $\vect{\mu}_q$ are modeled using neural networks with two hidden layers consisting of 512 units each, and the Exponential Linear Unit (ELU) function as the activation function.

Similarly, we model the policy as a Gaussian distribution 
\begin{align}
    \policy(\act|\stt,\latent) \sim \mathcal{N}(\vect{\mu}_{\pi}(\stt, \latent;{\theta}_{\pi}),\sigma_{\pi}^2\vect{I})
\end{align}
where $\theta_{\pi}$ denotes the neural network parameters. We adopt a mixture-of-expert (MoE) structure consisting of six expert networks, each of which has three hidden layers with 512 units with ELU as activation function. The parameters of these experts are combined based on weights calculated by a gating network that includes two hidden layers of 64 units each. The standard deviation of policy distribution $\sigma_{\pi}$ is set to 0.05. 

The world model $\omega(\stt, \act;\theta_w)$ is formulated as a deterministic neural network. It consists of four hidden layers, each with 512 units, and uses ELU activation functions. All of its parameters are collectively represented by $\theta_w$. The world model outputs both the skeleton state and the fatigue state, with the latter representing a prediction of the fatigue state for the next time step. The handling of the skeleton state is similar to the methods described in \cite{controlvae2022,supertrack2021}. Notably, this world model is formulated in maximal coordinates. During the early stages of training, the model can sometimes produce inaccurate bone positions. Such inaccuracies often result in excessive muscle lengths, leading to significant passive muscle forces and causing unstable training. To mitigate this, we employ a differentiable forward kinematics procedure, leveraging the predicted local rotation to prevent infeasible bone positions.

\subsection{Training}
\input{Sections/supp_training_controlVAE.tex}

\subsection{High-Level Policy}
Following \cite{controlvae2022}, we formulate the task policy $\policy(\latent^t|\stt^t,\task^t)$ as a Gaussian distribution $\mathcal{N}(\hat{\vect{\mu}}_{g},\sigma_{g}^2\vect{I})$ with a diagonal covariance $\sigma_{g}=\sigma_q$ and the mean function computed as 
\begin{align}
    \hat{\vect{\mu}}_{g}=\vect{\mu}_p + \vect{\mu}_{g}(\stt^{t}, \task^t;{\theta}_g)
\end{align}
where ${\theta}_g$ denotes the network parameters. We use a neural network with three hidden layers, each having 256 units, to model $\vect{\mu}_{g}$. The pseudocode for the training this task policy is outlined in Algorithm~\ref{alg:training_high_level}. The parameters $N_{\eqword{HL}}=512$ and $T_{\eqword{HL}}=16$ in our implementation.

\begin{algorithm}
  \SetAlgoLined    
  \DontPrintSemicolon
  
\SetKwProg{TrainVelocityControl}{Function}{:}{end}
\TrainVelocityControl{ \textnormal{\textbf{TrainVelocityControl}( $p$, $\world$,  $\pi$ ) } }
{ 
  Initialize $\simBuffer$ with random simulated trajectories $\{\tau\}$ \;
  $\loss_{\task} \gets 0$\;
  \For{$i\gets0$ \KwTo $N_{\textnormal{NL}}$}{
    Select a random task $\hat{\task}$ \;
    Sample ${\stt}^0$ from $\simBuffer$ \;
    \For{$t\gets0$ \KwTo $T_{\textnormal{NL}}$}{
      Compute task parameter $\task^t$ according to $\stt^t$ and $\hat{\task}$\;
      Sample $\latent \sim{} \policy_{\task}(\latent|\stt^t, \task^t)$ \;
      Sample $\act^t \sim{} \policy(\act^t|{\stt}^t,\latent^t)$ \;
      ${\stt}^{t+1} \leftarrow$ $\world({\stt}^t,\act^t)$ \;            
      $t \leftarrow t+1$\;

    }
    ${\tau}_{\task} \gets \{\stt^t, \latent^t \} $ \;
    $\loss_{\task} \gets \loss_{\task} + \loss_{\task}({\tau}_{\task})$ \;
  }      
  Update $\policy_{\task}$\ with $\loss_{\task}$\;
}     
\caption{Train High-Level Policy}
\label{alg:training_high_level}
\end{algorithm}

The loss functions have the form
\begin{align}
    \mathcal{L}(\tau_{\task})=
    \sum_{t=1}^{T} \left[\mathcal{L}_{\task}(\stt^{t}) + \mathcal{L}_{\eqword{fall}}(\stt^t) \right]
    + w_{z}\sum_{t=0}^{T-1} \Vert \vect{\mu}_{g} \Vert_2^2 ,
    \label{eqn:high_level_policy_loss}
\end{align}
where $\mathcal{L}_{\task}(\stt^{t})$ is the task-specific objective function. The $\mathcal{L}_{\eqword{fall}}$ term penalizes falling down. The regularization term ensures that the mean value shift between the goal prior and the fixed low-level prior remains low, ensuring the minimal change in motion quality. 

We use the heading control task in \cite{controlvae2022} to test \mvae{}.
In this task, the character is required to move in a specific direction indicated by the target direction $\theta_h\in{}[-\pi,\pi]$ at a given speed of $v\in{}[0.0,3.0]\,m/s$. The objective function for this task is defined based on the character's accuracy to reach its target direction while maintaining the specified speed. Specifically,
\begin{align}
    \mathcal{L}_{\task}(\stt) &= w_{\theta_h} | \theta_h^* - \theta_h | + w_{v} \frac{|v^*- v|}{\max(v^*, 1)},
    \label{eqn:heading_loss}
\end{align}
where $\theta_h$ and $v$ are the character's current heading direction and velocity, respectively. $\theta_h^*$ is the target heading direction and $v^*$  is the target velocity. $w_{\theta_h}=2.0$ and $w_{v}=1.0$ are balancing weights.

\subsection{Other Implementation Details}
\label{sec:implementation_details}

\paragraph{Fatigue State} 
The fatigue state of the character is characterized by the values of $M_\text{A}$, $M_\text{F}$, and $M_\text{R}$ for all the muscles. However, naively stacking all these variables into a single vector would lead to a very high-dimensional representation. To address this, we employ a more compact representation. We categorize all the muscles into five parts corresponding to the trunk and the four limbs. The fatigue state of the character, $\stt_{\text{fatigue}}$, is then defined by the average value of $M_\text{A}$, $M_\text{F}$, and $M_\text{R}$ for these five parts. We use a weighted average strategy to compute these values. In this approach, the fatigue parameters of each muscle are weighted by their maximum isometric force, $f_{\text{m0}}$. Since $f_{\text{m0}}$ is typically larger for major muscles, this strategy ensures that the fatigue states of the major muscles have a greater impact on the policy.

\paragraph{Fatigue Initialization}
During training, we initialize the fatigue variables $M_\text{A}$, $M_\text{F}$, and $M_\text{R}$ randomly each time the environment resets. To ensure a valid combination of these variables, we select a random point from a predetermined fatigue evolution curve, which is generated by tracking a synthetic target load pattern. A typical curve for this initialization is illustrated in \fig\ref{fig:3ccr_demo}.

\section{Experiments} 
\subsection{Character} 
\begin{figure}
  \includegraphics[width=0.45\textwidth]{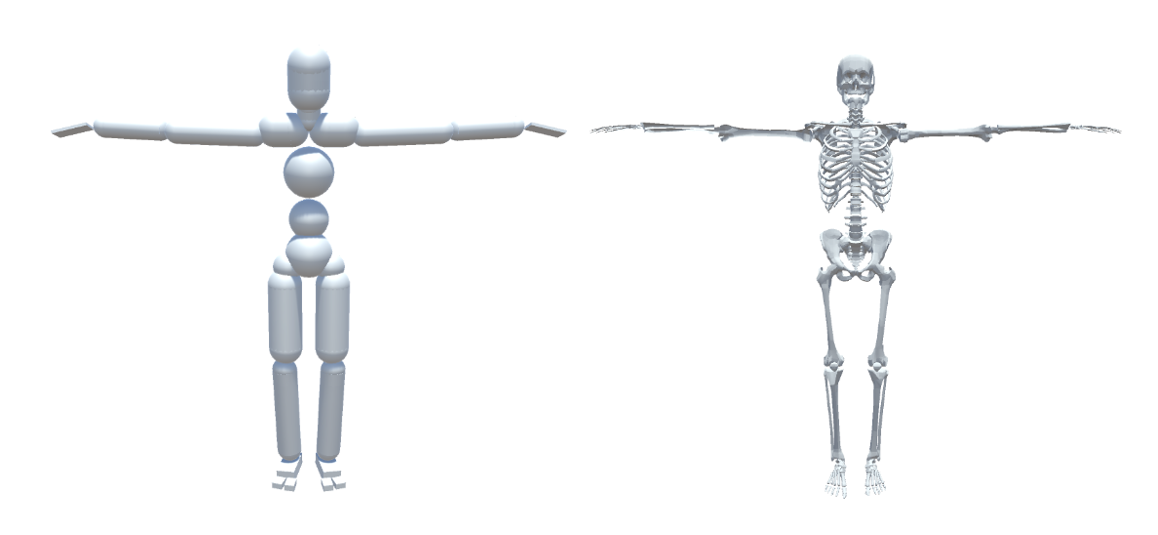}
  \caption{The physics collision geometries (left) and rendering mesh (right) of our character. The physics-based character is composed of 23 rigid bodies interconnected by 22 joints. We set the elbows and knees as hinge joints, while the other joints are set as ball-socket joints.  }
  \label{fig:character}
  \Description{}
\end{figure}

The character model depicted in Figure \ref{fig:character} is used in all our experiments. It has a height of $1.68$\,m, weighs $61.4$\,kg, consists of 23 rigid bodies connected by 22 joints, and is actuated by 284 muscles. The muscle model, including both the muscle dynamics and fatigue dynamics, operates at a frequency of $120$\,Hz. We implement the implicit joint damping mechanism to ensure the numerical stability of the simulation with a large timestep. The damping coefficient $k_{\text{d-joint}} = 10.0$ is applied uniformly to all joints. For each muscle, the stiffness parameter $k_{\text{p}}$  is set to the same value as the Hill-type maximum isotropic force, and the damping coefficient $k_{\text{d}}$ is set to $0.1 k_{\text{p}}$.  The original 3CC-model paper \cite{xia2008theoretical} suggests that there are three types of muscles: slow (S), fatigue-resistant (FR), and fast fatigue (FF). As a simplified model, we assume all the muscles are S-muscles. The parameters of fatigue are then ${F}=0.01$, ${R}=0.002$, $L_\text{D}=L_\text{R}=50.0$ and ${r}=2.0$. We also test other fatigue ratio in the experiment showed at the last paragraph of Section \ref{para:fatigue_exp}. We keep the ratio of $F$ over $R$ at $5$ for all experiments except the arm holding experiment, where $F/R=20, F=0.1$ for faster reaching the powerless posture of the arms.

\subsection{Dataset}
\begin{table}[t]
  \centering
  \caption{Motions Used for the Locomotion \mvae{}}
  \label{tab:motionlength}
  \begin{tabular}{|c|c|}
  \hline
  Motion & Frames (20\,fps) \\ \hline
  Walk  & 5227 \\ \hline
  Run   & 4757\\ \hline
  Jump  &  4889\\ \hline
  Run2(Test)  &  5477\\ \hline
  \end{tabular}
\end{table}
\Tab\ref{tab:motionlength} lists the motions used in our experiments. All these motions are selected from the LaFAN dataset \cite{harvey2020lafan}. The last row of \Tab \ref{tab:motionlength}  denotes the unseen motion clip of 8th demo in the supplementary video which is only used in testing rather than training. We use the dance motion from \cite{jehee19muscle} for testing which is the 9th demo in the supplementary video. The motion data of \emph{Jump Spin Kick} and \emph{Horse Stance} has already been mentioned in the main text.

\subsection{Muscle Render}
To more clearly reflect the muscle activation state, we use a linear relationship from white to red, with red indicating muscles that are more activated. Simultaneously, we increase the width of the muscle polylines in our visualization for muscles with higher activation levels.

%% file: Sections/supp_training_controlVAE.tex
\begin{algorithm}
    \SetAlgoLined    
    \DontPrintSemicolon
    
      \SetKwProg{Train}{Function}{:}{end}
      \Train{ \textnormal{\textbf{Train}$($ $)$  }}
      {
        Initialize $q$, $p$, $\pi$, $\world$, $\simBuffer{} \gets \varnothing  $ \;
        \While{\textnormal{not terminated}}{
            \tcp{collect simulation trajectories}
            Remove the oldest $N_{B}'$ simulation tuples from $\simBuffer{}$ \;
            \While{$|\simBuffer| < N_{B}$}{
                Select $\tilde{\tau}=\{\tilde{\stt}^0_{\text{skeleton}},\dots{},\tilde{\stt}^T_{\text{skeleton}}\}$ from \dataset \;
                ${\stt}^0 \leftarrow [\tilde{\stt}^0_{\text{skeleton}},\text{random}(\stt_{\text{fatigue}})]$\;
                ${\tau} \gets$ \textbf{GenerateTrajectory}($\tilde{\tau}$, ${\stt}^0$, $q$,  $\pi$, {None}, $|\tilde{\tau}|$) \;
                Store ${\tau}$ and $\tilde{\tau}$ in $\simBuffer$\;
            }
            \textbf{TrainWorldModel}($\world$, $T_w$, $\simBuffer$) \;
            \textbf{TrainMuscleVAE}($\world$, $q$, $p$, $\pi$, $T_{\text{VAE}}$, \simBuffer{} ) \;
        }
      }

      \SetKwProg{GenerateTrajectory}{Function}{:}{end}
      \GenerateTrajectory{ \textnormal{\textbf{GenerateTrajectory}$($ $\tilde{\tau}$, $\stt^0$, $q$, $\pi$, $\world$, $T$ $)$} }
      {    
        $t \gets 0$\;
        \While{\textnormal{not terminated $and$ $t < T$}}{
            \eIf { $\policy$ is \textnormal{a list} }
            {
                Extract $\act^t$ from $\policy$ \; 
            }
            {
            Extract $\tilde{\stt}^{t+1}_{\text{skeleton}}$ from $\tilde{\tau}$ \;
            Sample $\latent^t \sim{} q(\latent^t|{\stt}^t,\tilde{\stt}^{t+1}_{\text{skeleton}})$\;
            Sample $\act^t \sim{} \policy(\act^t|{\stt}^t,\latent^t)$ \;
            }
            
            ${\stt}^{t+1} \leftarrow$ Simulate$({\stt}^t$, $\act^t)$ \textbf{if} $\world$ is \textnormal{None} \textbf{else} $\world({\stt}^t,\act^t)$ \;
            
            $t \leftarrow t+1$\;
        }
        ${\tau}=\{{\stt}^0, \act^0, {\stt}^1, \act^1, \dots, \}$\; 
      }
      
      \SetKwProg{TrainWorldModel}{Function}{:}{end}
      \TrainWorldModel{ \textnormal{\textbf{TrainWorldModel}$($ $\world$, $T$, $\simBuffer{}$ $)$  }}
      {
        $\mathcal{L} \gets 0$\;
        \For{$i\gets0$ \KwTo $N_{\textnormal{batch}}$}{
            Sample ${\tau}^*=\{{\stt}^0, {\act}^0, {\stt}^1, {\act}^1, \dots, \}$ from $\simBuffer$, ignore $\tilde{\tau}^*$ \;
            $\pi^*\gets{}\{{\act}^0, {\act}^1, {\act}^2, \dots\}$ \;
            $\bar{\tau} \gets$ \textbf{GenerateTrajectory}({None}, ${\stt}^0$, {None}, $\pi^*$, $\world$, $T$) \;

            $\loss \gets \loss + \loss_{\text{w}}(\bar{\tau},{\tau}^*) $ \;
        
        }
        Update $\world$ with $\loss{}_{\eqword{w}}$
      }

      \SetKwProg{TrainMuscleVAE}{Function}{:}{end}
      \TrainMuscleVAE{ \textnormal{\textbf{TrainMuscleVAE}$($ $\world$, $q$, $p$, $\pi$, $T$, \simBuffer{} $)$} }
      { 
        $\mathcal{L} \gets 0$\;
        \For{$i\gets0$ \KwTo $N_{\textnormal{batch}}$}{
            Sample ${\tau}^*$ and $\tilde{\tau}^*$ from $\simBuffer$\;
            Extract  ${\stt}^0$ from ${\tau}^*$ \;
            ${\tau} \gets$ \textbf{GenerateTrajectory}($\tilde{\tau}^*$, ${\stt}^0$, $q$, $\pi$, $\world$, $T$) \;
            $\loss \gets \loss + \loss_{\text{rec}}({\tau},\tilde{\tau}^*)  + \beta \loss_{kl}({\tau}) +  \loss_{\text{act}}({\tau})$ \;
        }
        
        Update $q$, $p$, $\pi$ with $\loss$
      }     
      
    \caption{Train MuscleVAE}
    \label{alg:training_muscle_VAE}
\end{algorithm}

We employ the training algorithm from ControlVAE \cite{controlvae2022} to train our MuscleVAE model. In Brief, the training objective is to train the posterior distribution $q(\latent|\stt, \tilde{\stt}_{\text{skeleton}})$ and the policy $\policy(\act|\stt, \latent)$ to make the distribution of the generated motions $p(\tau)$ matches the distribution of a motion dataset $\dataset=\{\tilde{\tau}_i\}$. Here, the trajectory $\tau$ consists of a sequence of state $\{\stt^t\}$ and, if available, the correspond action $\{\act^t\}$. Algorithm~\ref{alg:training_muscle_VAE} outlines the major procedures of this algorithm. In this algorithm, $\simBuffer{}$ represents a buffer of simulation tuples, with each tuple consisting of a simulation state and its corresponding action. The parameters used in this training algorithm are set as follows: $N_B=5\times{}10^4$, $N'_{B}=2048$, $T_{\eqword{w}}=8$, $T_{\eqword{VAE}}=24$, and $N_{\eqword{batch}}=512$.